\documentclass[12pt,chicago,a4paper]{article}

\usepackage{amsmath,amsthm,amssymb,amscd,enumerate, natbib}
\usepackage[english]{babel}
\usepackage[latin1]{inputenc}
\usepackage{tikz}
\usepackage{listings}
\usepackage{graphicx,graphics}
\usepackage{bbm}
\usepackage{tocbibind}
\usepackage{fancyhdr}
\usepackage{pgf}

\usepackage{epsfig}
\usepackage{graphicx}              % grafica
\usepackage{listings}
\include{pdfpage}

\newcommand{\tc}{\textcolor}

\begin{document}
\title{A food chain ecoepidemic model:
infection at the bottom trophic level}
\author{Alessandra De Rossi, Francesco Lisa, Luca Rubini,\\
Alberto Zappavigna, Ezio Venturino\\
Dipartimento di Matematica ``Giuseppe Peano'',\\
Universit\`a di Torino, \\via Carlo Alberto 10, 10123 Torino, Italy,\\
email: ezio.venturino@unito.it}
\date{}

\maketitle

\begin{abstract}
In this paper we consider a three level food web subject to a disease affecting
the bottom prey. The resulting dynamics is much richer with respect to the purely demographic model,
in that it contains more transcritical bifurcations, gluing together the various equilibria, as well as
persistent limit cycles, which are shown to be absent in the classical case. Finally, bistability is discovered
among some equilibria, leading to situations in which the computation of their basins of attraction is relevant for
the system outcome in terms of its biological implications.
\end{abstract}

\textbf{AMS MR classification} 92D30, 92D25, 92D40

\section{Introduction}

Food webs play a very important role in ecology. Their study dates back to many years ago,
see \citet{[8],[9],[12],[15]}. The interest has not faded in time, since also recent
contributions can be ascribed to this field in mathematical biology, \citep{DA}, invoking the
use of network theory for managing natural resources, to keep on harvesting economic resources in
a viable way, without harming the ecosystems properties. This is suggested in particular in
the exploitation of aquatic environments.

Mathematical epidemiological investigations turned from the classical models \citep{H00} into studies encompassing
population demographic aspects about a quarter of a century ago \citep{BvdD,GH,MLH}. This step
allowed then,
on the other hand, the considerations of
models of diseases spreading among interacting populations \citep{HF}.
On the basic demographic structure of the Lotka-Volterra
model several cases are examined in \citet{V94}, in which the disease affects the prey as
well as the predators. In a different context, namely the aquatic environment,
diseases caused by viruses have been considered in \citet{BC}. More refined demographic
predator-prey models encompassing diseases affecting the prey have been proposed and
investigated in \citet{V95,CA99,AAMC}, while the case of infected predators has also been
considered \citep{V02,HV07}. In addition, other population associations such as competing
and symbiotic environments could host epidemics as well \citep{V01,V07,HV09,SMV10}.
It is worthy to mention that one very recent interesting paper reformulates intraguild
predator-prey models into an equivalent food web, when the prey are seen to be similar from the predator's
point of view \citep{SH}.
For a more complete introduction to this research field, see Chapter 7 of \citet{MPV}.

In \citet{DL}, the role of parasites in ecological webs
is recognized, and their critical role in shaping communities of populations is emphasized.
When parasites are accounted for, the standard pyramidal structure of a web gets almost
reversed, emphasizing the impact parasitic agents have on their hosts and in holding tightly together the web.
The role of diseases, in general, cannot be neglected, because, quoting directly from \citet{DA},
``Given that parasitism is the most ubiquitous consumer strategy, most food webs are
probably grossly inadequate representations of natural communities''.
%{\tc{red}{
A wealth of further examples
in this situation is discussed in the very recent paper \citet{SRH}.
%}}

Based on these considerations, then, in this paper we want to consider epidemics in a larger ecosystem,
namely a food system composed of three trophic levels. We assume that the disease affects only the prey at the
lowest level in the chain.

The paper is organized as follows. In the next Section we present the model, and its disease-free counterpart.
Section 3 contains the analytical results on the system's equilbria. A final discussion concludes the paper.

\section{The model}

We investigate a three level food web, with a top predator indicated by $W$, the intermediate population $V$ and the bottom prey $N$ that is affected by
an epidemic. It is subdivided into the two subpopulations of susceptibles $S$ and infected $I$. The disease, spreading by contact at rate $\beta$,
is confined to the bottom prey population.
We assume that neither one of the other populations can become infected by interaction with the infected prey.
The disease can be overcome, so that infected return to class $S$ at rate $\gamma$.
The top predators $W$ rely only on the intermediate population $V$ for feeding. They experience a natural mortality rate $m$, while they
convert captured prey into newborns at rate $p<h$, where the latter denotes instead their hunting rate on the lower trophic level $V$.
The gain obtained by the intermediate population
from hunting of susceptibles is denoted by $e$,
which must clearly be smaller than the damage inflicted to the susceptibles $c$, i.e.
$e<c$,
the corresponding loss rate of infected individuals in the lowest trophic level
due to capture by the intermediate population is $n$,
while $q<n$ denotes the return obtained by $V$
from capturing infected prey. The natural mortality
rate for the second trophic level is $l$. The natural plus disease-related mortality
for the bottom prey is $\nu$.
In this lowest trophic level, only the healthy prey reproduce,
at net rate $a$
and the prey environment carrying capacity is $K$.
The remaining terms in
the last equation indicate hunting losses and disease dynamics as mentioned above.
The model is then given by the following set of equations
\begin{eqnarray}\label{sistema}
\begin{split}
\frac{dW}{dt}&=-m W + p VW \\
\frac{dV}{dt}&=-lV + eSV - hVW + qIV \\
\frac{dI}{dt}&=\beta IS - n IV - \gamma I - \nu I \\
\frac{dS}{dt}&=aS\left(1 -\frac{S+I}{K}\right) - cVS - \beta SI +\gamma I 
\end{split}
\end{eqnarray}

The Jacobian of (\ref{sistema}) is
\begin{equation}\label{jacobian}
J=
\left[
\begin{array}{cccc}
-m+pV & pW & 0 & 0\\
-hV & -l+eS-hW+qI & qV & eV \\
0 & -n I & \beta S-nV-\gamma -\nu & \beta I \\
0 & -c S & -aS \frac 1K -\beta S+\gamma & J_{44}
\end{array}
\right]
\end{equation}
with
$$
J_{44}=a- \frac aK (2S+I) -cV-\beta I.
$$

%\newpage
\subsection{The model without disease}\label{sec:dis-free}
For later comparison purposes, we now discuss briefly the
food chain with no epidemics. It is obtained by merging the last two equations of (\ref{sistema}) and
replacing the two subpopulations of susceptibles and infected by the total prey
population $N=S+I$. We need also to drop the term containing $I$ in the second equation and replace $S$ by $N$ in it.
The last equation then becomes
$$
\frac{dN}{dt}=aN \left(1 -\frac{N}{K}\right) - cVN
$$
Correspondingly, the Jacobian $J$ changes into a $3\times 3$ matrix $\widehat J$,
by dropping the third row and column, dropping the terms in $I$ in $J_{22}$ and $J_{44}$
and again replacing with $N$ the population $S$.

In the $W-V-N$ phase space,
this disease-free model has four equilibria, the origin, which is unconditionally unstable, the bottom prey-only
equilibrium $Q_1=(0,0,K)$, the top-predator-free equilibrium
$\widetilde Q=\left( 0, \widetilde V, \widetilde N\right)$,
\begin{eqnarray*}
\widetilde V=\frac ac\left( 1-\frac l{eK}\right), \quad \widetilde N=\frac le
\end{eqnarray*}
and the coexistence equilibrium
$Q^*=\left( W^*, V^*, N^*\right)$, with
\begin{eqnarray*}
W^*=\frac 1h\left[ eK \left( 1- \frac {cm}{ap}\right) - l\right], \quad
V^*=\frac mp, \quad
N^*=K \left( 1- \frac{cm}{ap}\right).
\end{eqnarray*}

Now, $Q_1$ is stable if
\begin{equation}\label{Q1_stab}
1<\frac l{eK},
\end{equation}
and furthermore the eigenvalues of the Jacobian are all real, so that no Hopf bifurcation can arise at this point.
Instead $\widetilde Q$ is feasible if the converse condition holds,
\begin{equation}\label{Qtilde_feas}
1\ge \frac l{eK},
\end{equation}
indicating a transcritical bifurcation between $Q_1$ and $\widetilde Q$. Stability for $\widetilde Q$ holds
when its first eigenvalue is negative, i.e. for
\begin{equation}\label{Qtilde_stab}
\frac mp >  \frac ac\left( 1-\frac l{eK}\right).
\end{equation}
Again here no Hopf bifurcations can arise, since the trace of the remaining $2\times 2$ submatrix is always strictly positive.
Feasibility for $Q^*$ is given by instead by the opposite of the above condition
\begin{equation}\label{Q*_feas}
\frac mp \le  \frac ac\left( 1-\frac l{eK}\right),
\end{equation}
indicating once more a transcritical bifurcation for which $Q^*$ emanates from $\widetilde Q$. Since two of the Routh-Hurwitz
conditions hold easily,
\begin{equation}\label{Q*_RH_1_3}
-{\textrm{tr}}\left(\widehat J(Q^*)\right)= \frac aK N^*>0, \quad -\det\left(\widehat J(Q^*)\right)= \frac aK l p W^*V^*N^*>0
\end{equation}
stability of $Q^*$ depends on the third one,
\begin{equation}\label{Q*_RH_2}
\frac aK N^* \left( pl W^* + ec N^* \right) V^* > \frac aK N^*  l p V^* W^*,
\end{equation}
which becomes $pl W^* + ec N^* > l p W^*$. The latter
clearly holds unconditionally. Therefore $Q^*$ is always locally asymptotically stable, and in view that no other equilibrium
exists when $Q^*$ is feasible, nor Hopf bifurcations from (\ref{Q*_RH_2}) are seen to arise, it is also globally asymptotically stable.
The same result holds for the remaining two equilibria
$Q_1$ and $\widetilde Q$ whenever they are locally asymptotically stable.
In fact, the three points $Q_1$, $\widetilde Q$ and $Q^*$ are in pairs mutually exclusive,
i.e. the equilibria $Q_1$, $\widetilde Q$ 
cannot be both simultaneously feasible and stable
and similarly for $\widetilde Q$, $Q^*$, in view
of the transcritical bifurcations that exist among them.

The global stability result for these equilibria can also be established
analytically with the use of
a classical method. In each case a Lyapunov function
can be explicitly constructed based on
the considerations of e.g. \citet{HS}, p. 63. We have the following results.

\vspace{0.3cm}
%\newpage
{\textbf{Proposition 1.}}
Whenever equilibrium $Q_1$ is locally asymptotically stable, it is also globally asymptotically stable, using
$$
L_1=\frac {ch}{ep} W + \frac ce V + \left( N -K \ln \frac NK\right).
$$

\textbf{Proof.}
In fact, upon differentiating along the trajectories and simplifying, we find
$$
L_1'=-m \frac {ch}{ep} W + c \left( K-\frac le \right) V - \frac aK \left( N -K \right)^2,
$$
which is negative exactly when (\ref{Q1_stab}) holds.

\vspace{0.3cm}
{\textbf{Proposition 2.}}
Whenever equilibrium $\widetilde Q$ is locally asymptotically stable, it is also globally asymptotically stable, using
\begin{eqnarray*}
\widetilde L=\frac {ch}{ep} W + \frac ce \left( V - \widetilde V \ln \frac V{\widetilde V } \right)+
\left( N - \widetilde N \frac \ln N{\widetilde N} \right).
\end{eqnarray*}

\textbf{Proof.}
Here differentiation along the trajectories leads to
$$
\widetilde L'=\frac {ch}e \left( \widetilde V -\frac mp \right) W
- \frac aK \left( N - \frac le \right)^2,
$$
again negative exactly when (\ref{Qtilde_stab}) holds.

\vspace{0.3cm}
{\textbf{Proposition 3.}} 
Whenever equilibrium $Q^*$ is locally asymptotically stable, it is also globally asymptotically stable, using
$$
L^*=\frac {ch}{ep} \left( W- W^* \ln \frac W{W^*}\right)+
\frac ce (V-V^* \ln \frac V{V^*})
+(N- N^* \ln \frac N{N^*}).
$$

\textbf{Proof.}
In this case the calculation of the derivative, taking into account the fact that the population values
at equilibrium satisfy the nontrivial algebraic equilibrium equations stemming from the right hand side
of the differential system, leads easily to $(L^*)'=-aK^{-1}(N-N^*)^2<0$.

\subsection{Two particular cases}
For completeness sake, here we also briefly discuss the SIS model with logistic growth
and the subsystem made of the two lowest trophic levels.

\subsubsection{SIS model with logistic growth}\label{SIS}

Focusing on the last two equations of (\ref{sistema}), in which $V$ is not present, we obtain
an epidemic system with demographics, of the type investigated for instance in \citet{GH,MLH}, see also
\citet{H00}. It admits only two equilibria, in addition to the origin, which is unconditionally unstable,
namely the points $P_1=(0,K)$ and $P^*=(I^*,S^*)$, with
$$
S^*=\frac {\gamma+\mu}{\beta}, \quad
I^*=aS^* \frac{K-S^*}{(a+\beta K)S^*-\gamma K}=aS^* \frac {\beta K -\gamma -\nu}{a(\gamma+\nu)+\beta K \nu}.
$$
The latter is feasible for
\begin{equation}\label{P*_feas}
\beta K > \gamma + \nu.
\end{equation}
Stability of $P_1$ is attained exactly when the above condition is reversed, thereby showing the
existence of a transcritical bifurcation for which $P^*$ originates from $P_1$ when the infected
establish themselves in the system. This identifies also the disease basic reproduction number
$$
R_0= \frac {\beta K} {\gamma + \nu}.
$$
Thus when $R_0>1$, the disease becomes endemic in the system. Denoting by $J^*$ the Jacobian of this SIS system
evaluated at $P^*$, we find that
$$
-{\textrm{tr}}\left(J^*\right)=\gamma \frac {I^*}{S^*} + \frac aK S^* >0, \quad
\det \left(J^*\right) = \frac {\beta}K I^* \left[ S^* \left( a + K\beta\right) -K\gamma \right].
$$
The quantity in the last bracket is always positive, since it reduces to $a(\gamma +\nu)+\beta K \nu>0$.
Hence the Routh-Hurwitz conditions hold, i.e. whenever feasible, the endemic equilibrium is always stable.
In case $\gamma=0$, using once again \citet{HS} p. 63, we have the following result.

\vspace{0.3cm}
{\textbf{Proposition 4.}} 
For the SI model global stability for both equilibria $P_1$ and $P^*$ holds, whenever they are feasible and stable.
\vspace{0.3cm}

\textbf{Proof.} 
The Lyapunov function in this case is given by
$$
L^{SI}_1= \frac {a+\beta K}{\beta K}I +(S-K\ln \frac SK), \quad
L^{SI}_*=\frac {a+\beta K}{\beta K}(I-I^*\ln \frac I{I^*}) +(S-S^*\ln \frac S{S^*}).
$$
Differentiating along the trajectories for the former we find 
$$
(L^{SI}_1)'=\frac {a+\beta K}{\beta K}I (K\beta -\nu)- \frac aK (S-K)^2
$$
and the first term is negative if $P_1$ is stable, i.e. when (\ref{P*_feas}) does not hold.
For the second one, the argument is straightforward, leading to
$(L^{SI}_*)'= -aK^{-1} (S-S^*)^2 <0$.
\vspace{0.3cm}

The global stability for the more general case of $\gamma \neq 0$ has been discussed in the classical
paper \citet{BeCa}. For more recent results, see \citet{Vargas}.

\subsubsection{Subsystem with only the two lowest trophic levels.}

The second particular case is a basic ecoepidemic model, of the type studied in
\citet{V95} and to which we refer the interested reader.
From the analytic side, we just remark the coexistence equilibrium attains the population values
$$
I^* =\frac 1q (l-eS^*), \quad V^*=\frac 1n [\beta S^* - ( \gamma + \nu)]
$$
where $S^*$ solves the quadratic $B_2 S^2+ B_1 S +B_0=0$, with
$$
B_2= \frac {ae}{qK}-\frac aK+\frac {\beta e}q-\frac {c\beta}n, \quad
B_1= a-\frac {al}{qK}+\frac cn(\gamma +\nu) -\frac {\beta l}q-\frac {e\gamma}q, \quad
B_0= \frac {\gamma l}q>0.
$$
Existence and feasibity of the coexistence equilibrium can be discussed on the basis of the signs
of these coefficients. The characteristic equation $\sum_{i=0}^3 A_i \Lambda^i=0$ arising 
from the Jacobian of the subsystem evaluated at the coexistence equilibrium, $J^*$,
has the following coefficients
\begin{eqnarray*}
A_2&=&-{\mathrm{tr}}(J^*)=\gamma {I^*}{(S^*)^{-1}}+ aK^{-1} S^*>0,\\
A_1&=&qn I^* V^*+ce S^* V^* -\frac {\beta}K I^* [(a+\beta K)S^* - \gamma K],\\
%-{\mathrm{tr}}(J^*)=\gamma \frac {I^*}{S^*}+\frac aK S^*>0, \quad
A_0&=&-\det (J^*)=cq\beta V^*I^*S^* +\frac{en}K I^*V^* [K\gamma -S^*(a+K \beta)] - A_0 nqI^*V^*.
%{\mathrm{tr}}(J^*).
\end{eqnarray*}
For stability, the Routh-Hurwitz conditions require
\begin{eqnarray}\label{ecoepid_st}
A_0>0, \quad A_1>0, \quad A_2>0, \quad A_2A_1>A_0.
\end{eqnarray}
We will further discuss this point later.

\section{Model Analysis}
\subsection{Boundedness}

We define the global population of the system as $\psi(t) = W + V + I + S$.
Recalling the assumptions on the parameters, for which $e < c$, $q<n$  and $p<h$, we obtain the following inequalities
\begin{eqnarray*}
\frac {d \psi}{dt} &=& aS \left( 1 -\frac {S+I}{K}\right) - cVS - \beta SI +\gamma I + \beta IS - n IV - \gamma I - \nu I -lV \\
& &+ eSV - hVW + qIV -m W + p VW = aS \left( 1 -\frac{S+I}{K}\right) \\
&& - (c-e) VS -(n-q)IV - (h-p)VW - lV - m W -\nu I \\
&&<  aS \left( 1 -\frac{S}{K}\right) - lV - m W -\nu I.
\end{eqnarray*}

Taking now a suitable constant $ 0 < \eta < \min (\nu, l, m)$ we can write
\begin{eqnarray*}
\frac{d \psi}{dt} + \eta \psi 
% & = aS - a\frac{S^2}{K} - lV - m W -\nu I + \eta S + \eta I + \eta V + \eta W =\\ & 
&&< (a+\eta) S - a\frac{S^2}{K} + (\eta - \nu) I + (\eta - l) V + (\eta - m) W \\
& &\le (a+\eta) S - a\frac{S^2}{K} \leq \frac{K(a + \eta)^2}{4a} = L_1.
\end{eqnarray*}
From the theory of differential inequalities we obtain an upper bound on the total environment population
\begin{equation*}
0 \leq \psi(t) < \frac{L_1}{\eta} (1-e^{-\eta t}) + \psi(0) e^{-\eta t},
\end{equation*}
from which letting $t \rightarrow + \infty$ it ultimately follows that $\psi(t) \rightarrow {L_1}\eta^{-1}$,
which means that the total population of the system,
and therefore each one of its subpopulations, is bounded by a suitable constant
$$
\psi(t) \le M := \max \left\{ \frac{L_1}{\eta}, \psi(0)\right\}.
$$

\subsection{Critical points}

$\mathbf{E}_1 \equiv (0,0,0,0)$ is a clearly feasible but unstable equilibrium, with eigenvalues $-m$, $-l$, $-\gamma - \nu$, $a$.

$\mathbf{E}_2 \equiv (0,0,0,K)$ is feasible and conditionally stable. This equilibrium coincides with $Q_1$ of the classical disease-free
system.
The Jacobian's eigenvalues at $\mathbf{E}_2$ are
$-m$, $-l+eK$, $\beta K -\gamma - \nu$, $-a$, so that stability is ensured by
\begin{eqnarray}\label{E2_stab}
K < \min \left\{ \frac{l}{e} ,\frac{\gamma + \nu}{\beta}\right \}.
\end{eqnarray}
Note that the stability of this equilibrium now hinges also on the epidemic parameters, so that this equilibrium may be stable in the
disease-free system, but may very well not be stable when a disease affects the population at the bottom trophic level in the ecosystem.

We have then the disease-free equilibrium with all the trophic levels,
$$
\mathbf{E}_3 \equiv \left(W_3,\frac{m}{p},  0, K\frac{a p-c m}{a p}\right), \quad W_3=\frac{a p K e-m e c K-a p l}{a h p}.
$$
feasible for
%with
\begin{equation}\label{E3_feas}
eK(ap - mc) \ge apl .
\end{equation}
Again we observe that since $W_3\equiv W^*$, the equilibrium $\mathbf{E}_3$ coincides with the coexistence equilibrium $Q^*$ of the disease-free food chain.
Thus also its feasibility condition (\ref{E3_feas}) reduces to (\ref{Q*_feas}).

One eigenvalue easily factors out, to give the stability condition
\begin{equation}\label{E3_stab}
a K p \beta< c K \beta m+a n m+a p \gamma+a p \nu.
\end{equation}
The reduced $3\times 3$ Jacobian $\widetilde J$ then gives a cubic characteristic equation, the Routh-Hurwitz conditions for which become
\begin{eqnarray*}
-\textrm{tr}\widetilde {J}(\mathbf{E}_3)=\frac 1p (ap-cm)>0, \quad
%\widetilde {\textrm{M}_2}(\mathbf{E}_3)=\widetilde {J_{12}}(\mathbf{E}_3) \frac {hm}p + cemK \frac {ap-cm}{ap^2}>0, \\
%-\det \widetilde {J}(\mathbf{E}_3)= \frac {hm}p \widetilde {J_{12}}(\mathbf{E}_3)(ap-cm)>0,
-\det \widetilde {J}(\mathbf{E}_3)= hm W_3 (ap-cm)>0,
\end{eqnarray*}
corresponding to (\ref{Q*_RH_1_3}).
In fact, these conditions hold in view of feasibility, (\ref{E3_feas}).
Also the third Routh-Hurwitz condition is the same as (\ref{Q*_RH_2}) so it is satisfied.

Stability of $\mathbf{E}_3$ hinges however also on the first eigenvalue, i.e. on
condition (\ref{E3_stab}). Again, the presence of the epidemics-related parameters in it, shows that the
behavior of the demographic ecosystem is
affected by the presence of the disease.

Next, we find the subsystem in which only the intermediate population and the bottom healthy prey thrive,
$$
\mathbf{E}_4 \equiv \left( 0, \frac{a (K e-l)}{e c K}, 0, \frac{l}{e}\right)
$$
which is feasible iff 
\begin{equation}\label{E4_feas}
K\ge \frac{l}{e}.
\end{equation}

The eigenvalues of the Jacobian are
\begin{eqnarray*}
\lambda_1 = \frac{-l a+\sqrt{l^2 a^2-4 l a e^2 K^2+4 l^2 a e K}}{2 e K} \\
\lambda_2 = \frac{-l a-\sqrt{l^2 a^2-4 l a e^2 K^2+4 l^2 a e K}}{2 e K}\\
\lambda_3 = \frac{-m e K c-a p l+a p e K}{e K c} \\
\lambda_4 = \frac{\beta l K c+a n l-a n e K-\gamma e K c-\nu e K c}{e K c}.
\end{eqnarray*}

We have a pair of complex conjugate eigenvalues with negative real part if and only if
$l a +4 l e K \leq 4 e^2 K^2$. In the opposite case, these eigenvalues are real: $\lambda_2 <0$ always, while $\lambda_1 <0$ holds if
$ l < eK$, i.e. ultimately in view of the feasibility condition (\ref{E4_feas}). In any case, the stability of this equilibrium is
always regulated by the remaining real eigenvalues.
Therefore the point $\mathbf{E}_4$ is stable if and only if these two conditions are verified
\begin{eqnarray}\label{E4_stab}
a p e K < m e K c+a p l, \quad
\beta l K c+a n l < a n e K+\gamma e K c+\nu e K c .
\end{eqnarray}

{\textbf{Remark 1.}} Note that the first above condition is the opposite of (\ref{E4_feas}),
or equivalently as remarked earlier, (\ref{Q*_feas}).
Hence there is a transcritical bifurcation between $\mathbf{E}_3$ and $\mathbf{E}_4$ which
has the classical counterpart $Q^*$ and $\widetilde Q$.

No Hopf bifurcations can arise at this point, since the real part of the first two eigenvalues cannot vanish.

The point at which just the bottom prey thrives, with endemic disease, is
$$
\mathbf{E}_5 \equiv \left( 0, 0, \frac{a (K \beta \gamma+K \beta \nu- \gamma^2-2 \gamma \nu-\nu^2)}
{\beta (a \gamma+a \nu+K \beta \nu)}, \frac{\gamma+\nu}{\beta}\right).
$$
It is feasible for
\begin{equation}\label{E5_feas}
\beta K \ge \gamma + \nu.
\end{equation}

{\textbf{Remark 2.}} Observe that the stability condition (\ref{E2_stab})
for $\mathbf{E}_2$ can fail in two different ways.
If $(\gamma+\nu) \beta^{-1}>le^{-1}$, then (\ref{E2_stab}) is the opposite condition of feasibility for $\mathbf{E}_4$, (\ref{E4_feas}).
This transcritical bifurcation corresponds thus to the one
between $Q_1$ and $\widetilde Q$ in the classical disease-free model.

{\textbf{Remark 3.}} On the other hand, for $(\gamma+\nu) \beta^{-1}<le^{-1}$,
we discover a transcritical bifurcation between $\mathbf{E}_2$ and $\mathbf{E}_5$, see (\ref{E2_stab}) and (\ref{E5_feas}).
This situation clearly does not exist in the classical model.

Since the characteristic equation factors,
two eigenvalues come from a $2\times 2$ minor $\widetilde J$ of the Jacobian, for which
\begin{eqnarray}\label{tr_E5}
-{\textrm{tr}} \left( \widetilde J(\mathbf{E}_5)\right) =
\frac a{\beta K}\frac{K \beta \nu^2+a \gamma^2+2 a \gamma \nu+a \nu^2+\beta^2 K^2 \gamma-\beta K \gamma^2}{a \gamma+a \nu+K \beta \nu}\\ \label{det_E5}
\det \left( \widetilde J(\mathbf{E}_5)\right) =
a \frac{a \gamma+a \nu+K \beta \nu}{\beta K}\ \
\frac{K \beta \gamma+K \beta \nu-\gamma^2-2 \gamma \nu-\nu^2}{a \gamma+a \nu+K \beta \nu}.
\end{eqnarray}
The Routh-Hurwitz conditions for stability then require positivity of both these quantities. Now (\ref{det_E5}) is implied by feasibility, (\ref{E4_feas}),
while (\ref{tr_E5}) yields
$$
K \beta \nu^2+a (\gamma+\nu)^2+\beta^2 K^2 \gamma>\beta K \gamma^2.
$$
But this condition always holds: note that the left hand side is minimized by taking $\nu=0$. The resulting inequality,
$$
\frac {a\gamma}{\beta K}+\beta K > \gamma
$$
holds because
the right hand side is larger than $\beta K$ and using (\ref{E5_feas}) the latter exceeds $\gamma+\nu\ge \gamma$.
The remaining two eigenvalues are immediate, the first one is $-m$, the other one gives the stability condition
\begin{equation}\label{E5_stab}
\frac{e (\gamma+\nu)}{\beta}+\frac{q a [K \beta \gamma+K \beta \nu-(\gamma+\nu)^2]}{\beta (a \gamma+a \nu+K \beta \nu)}<l.
\end{equation}

We find up to two equilibria in which the top predators disappear,
$$
\mathbf{E}_{6,7} \equiv \left( 0,\frac{\beta \widehat S -\gamma - \nu}{n},\frac{l-e \widehat S}{q}, \widehat S\right),
$$
where $\widehat S$ are the roots of the following quadratic equation:
\begin{equation}\label{eq:definitionOfS}
\widetilde {A} S^2 + \widetilde {B} S + \widetilde C = 0
\end{equation}
where
$\widetilde {A}:= n a q-n e a+\beta c K q-n e \beta K$,
$\widetilde {B}:=n l a+K \beta l n-q K c \gamma-c K \nu q+n \gamma K e-q K a n$,
$\widetilde {C}:=-n l \gamma K$.
Feasibility imposes the following requirement on the roots $\widehat S$,
\begin{equation}\label{E7_feas}
\frac{\gamma + \nu}{\beta}< \widehat S < \frac{l}{e},
\end{equation}
which imply their positivity. In turn, using Descartes' rule, since $\widetilde {C}<0$, the latter is a consequence of either one of the conditions
$$
\widetilde {A} > 0, \quad \widetilde {B} > 0 
$$
which lead respectively to either one of the following explicit conditions
$$
q(na+\beta c K) > ne(a+\beta K), \quad
n(la + K(\beta l + \gamma e)) > qK(an+c(\gamma + \nu)).
$$

\begin{figure}[htbp]
\centering
\title{Top-predator-free equilibrium.}\\
\begin{minipage}[c]{.40\textwidth}
\centering
\includegraphics[width=1\textwidth]{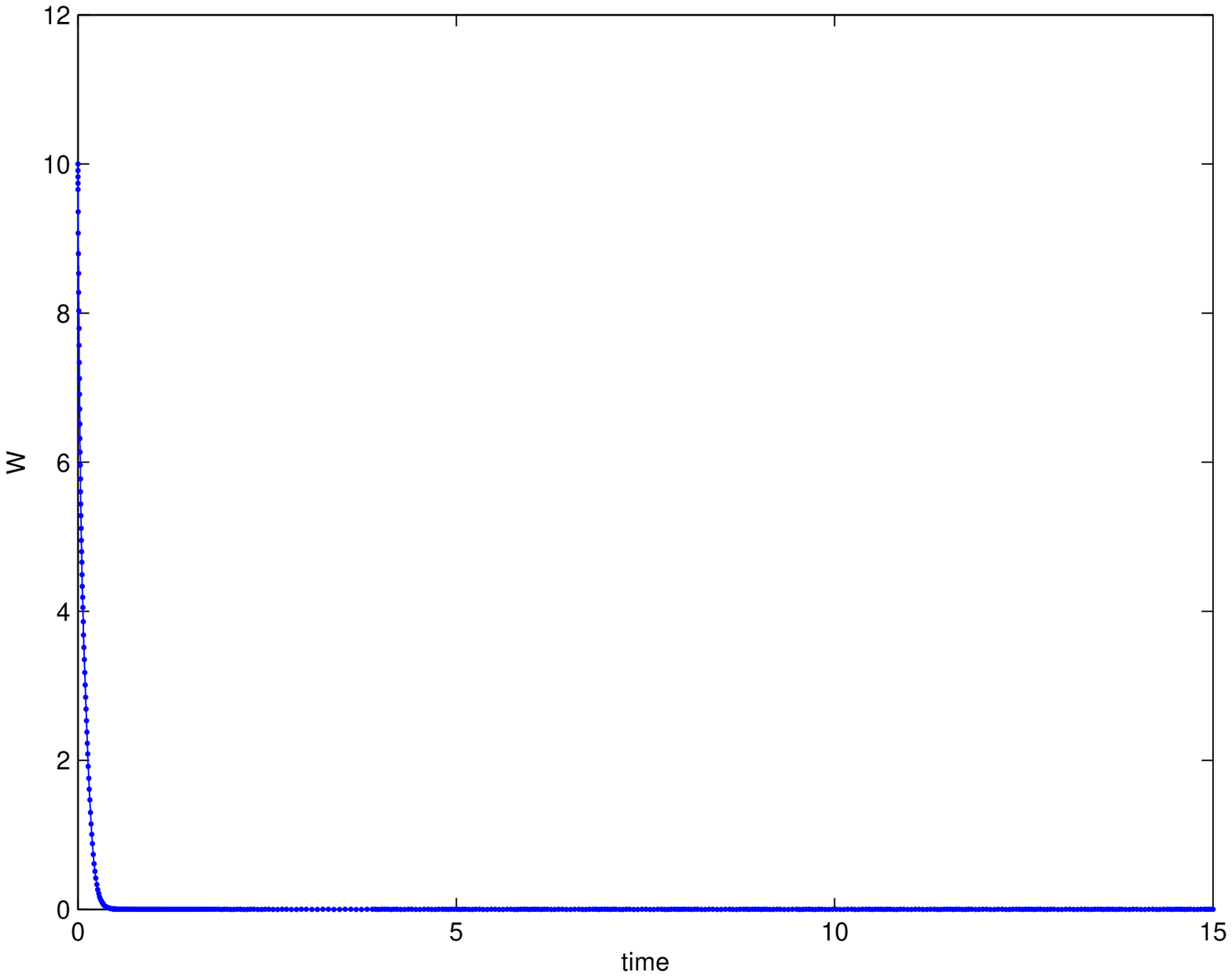}
\end{minipage}%
\hspace{10mm}%
\begin{minipage}[c]{.40\textwidth}
\centering
\includegraphics[width=1\textwidth]{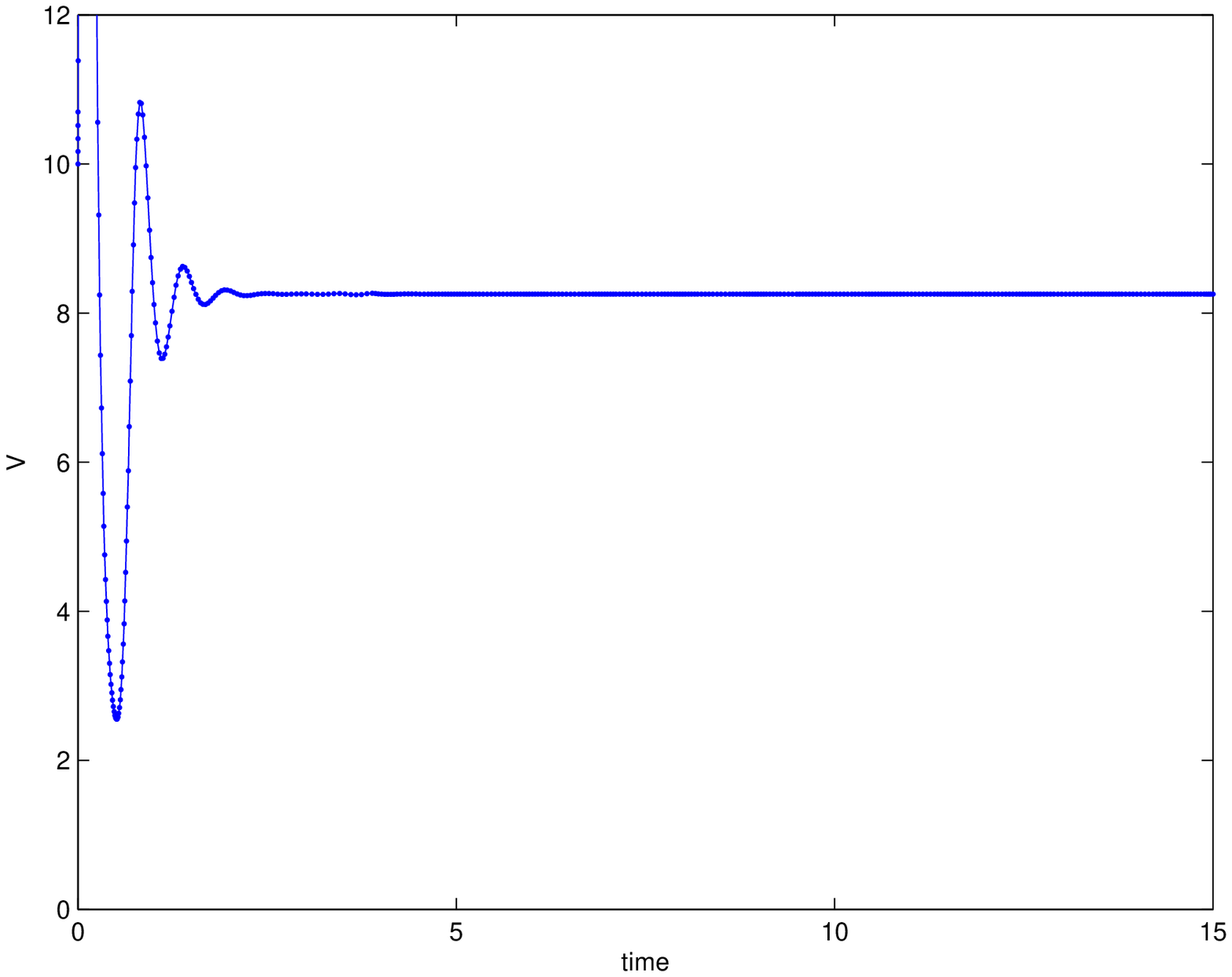}
\end{minipage}
\begin{minipage}[c]{.40\textwidth}
\centering
\includegraphics[width=1\textwidth]{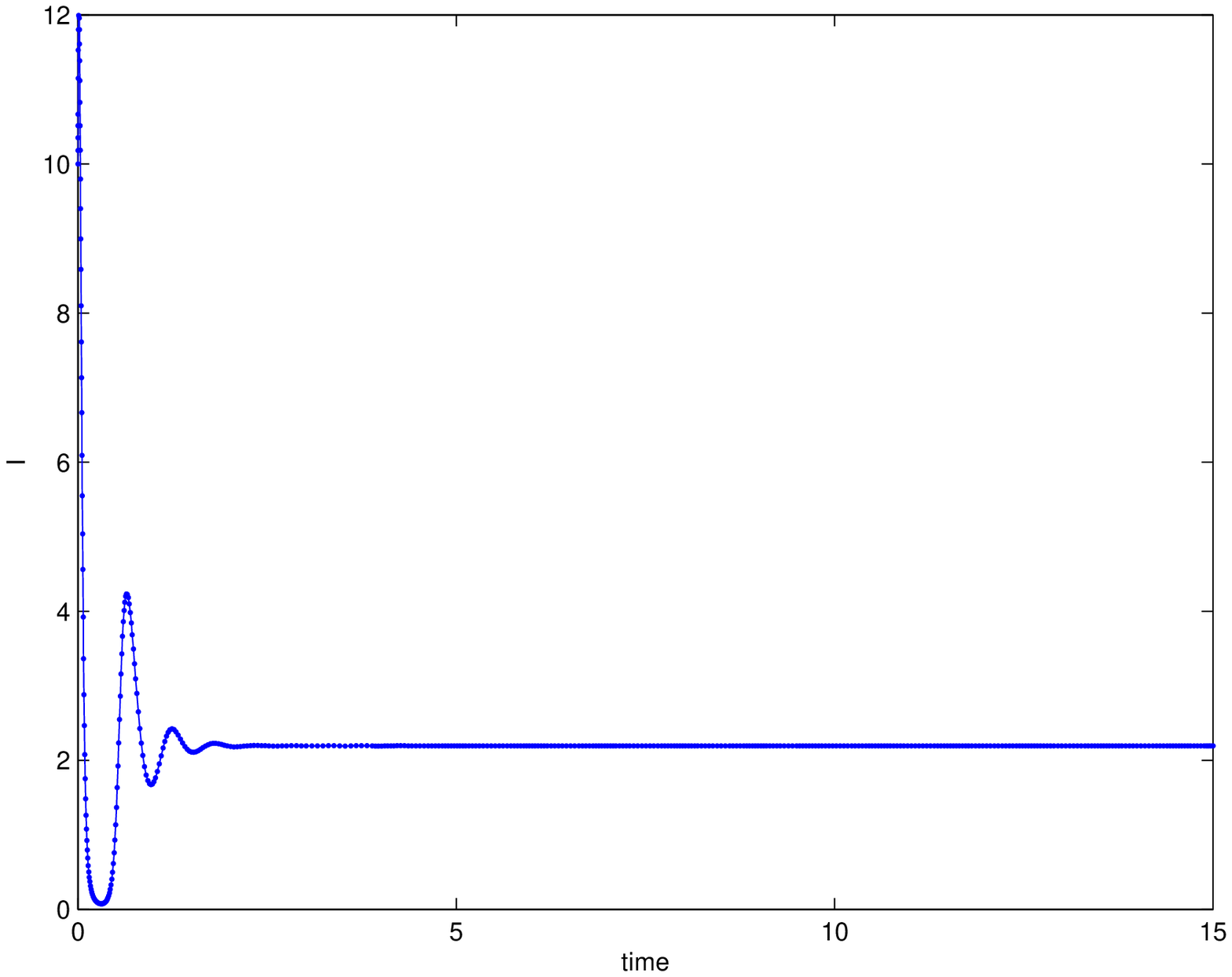}
\end{minipage}%
\hspace{10mm}%
\begin{minipage}[c]{.40\textwidth}
\centering
\includegraphics[width=1\textwidth]{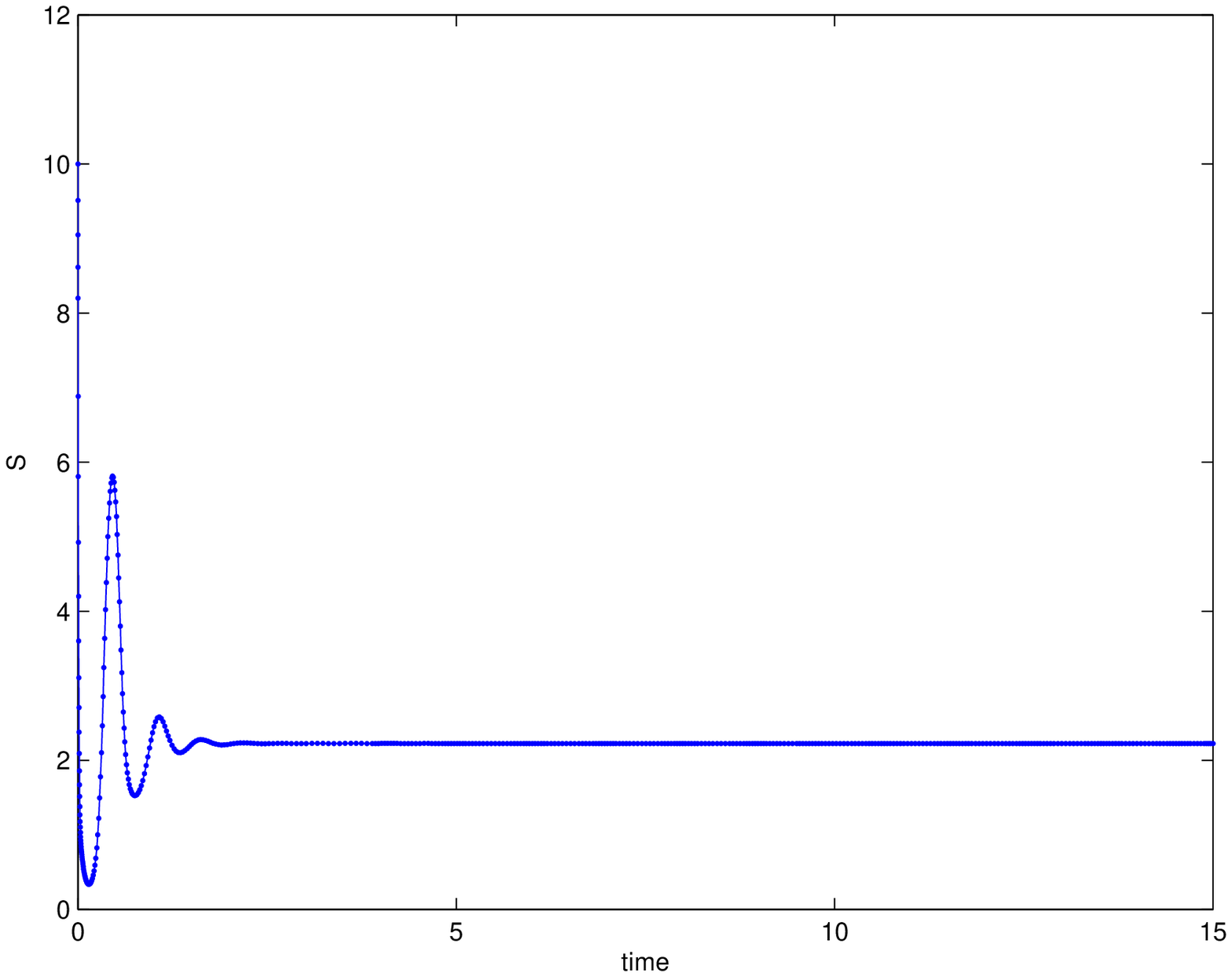}
\end{minipage}
\caption{Here and in all the following figures,
the plots are shown starting from the left top corner in clockwise order to represent the populations $W$, $V$, $S$, $I$
as function of times.
Here they are obtained by
the parameter set $a = 30$, $K=7.5$, $m = 20$, $p = 0.2$, $
l = 11$, $e = 1$, $h = 0.5$, $q = 4$, $\beta = 5$, $n = .5$, $c = 1$, $\gamma = 7$, $\nu=0$.}
\label{fig_a}
\end{figure}

One eigenvalue factors out, implying for stability an upper bound for the bottom healthy prey population,
$$
\widehat S<\frac {m n+p \gamma+p \nu}{p \nu }.
$$
The remaining characteristic equation is a complicated cubic. Therefore stability of this equilibrium is analysed only numerically.
By choosing the following hypothetical set of parameters, $a = 30$, $K = 7.5$, $m = 20$, $p = 0.2$,
$l = 11$, $e = 1$, $h = 0.5$, $q = 4$, $\beta = 5$, $n = .5$, $c = 1$, $\gamma = 7$, $\nu=0$, the system settles to this equilibrium, Figure \ref{fig_a}.

\begin{figure}[htbp]
\centering
\title{Hopf bifurcation at the top-predator-free equilibrium.}\\
\begin{minipage}[c]{.4\textwidth}
\centering
\includegraphics[width=.78\textwidth]{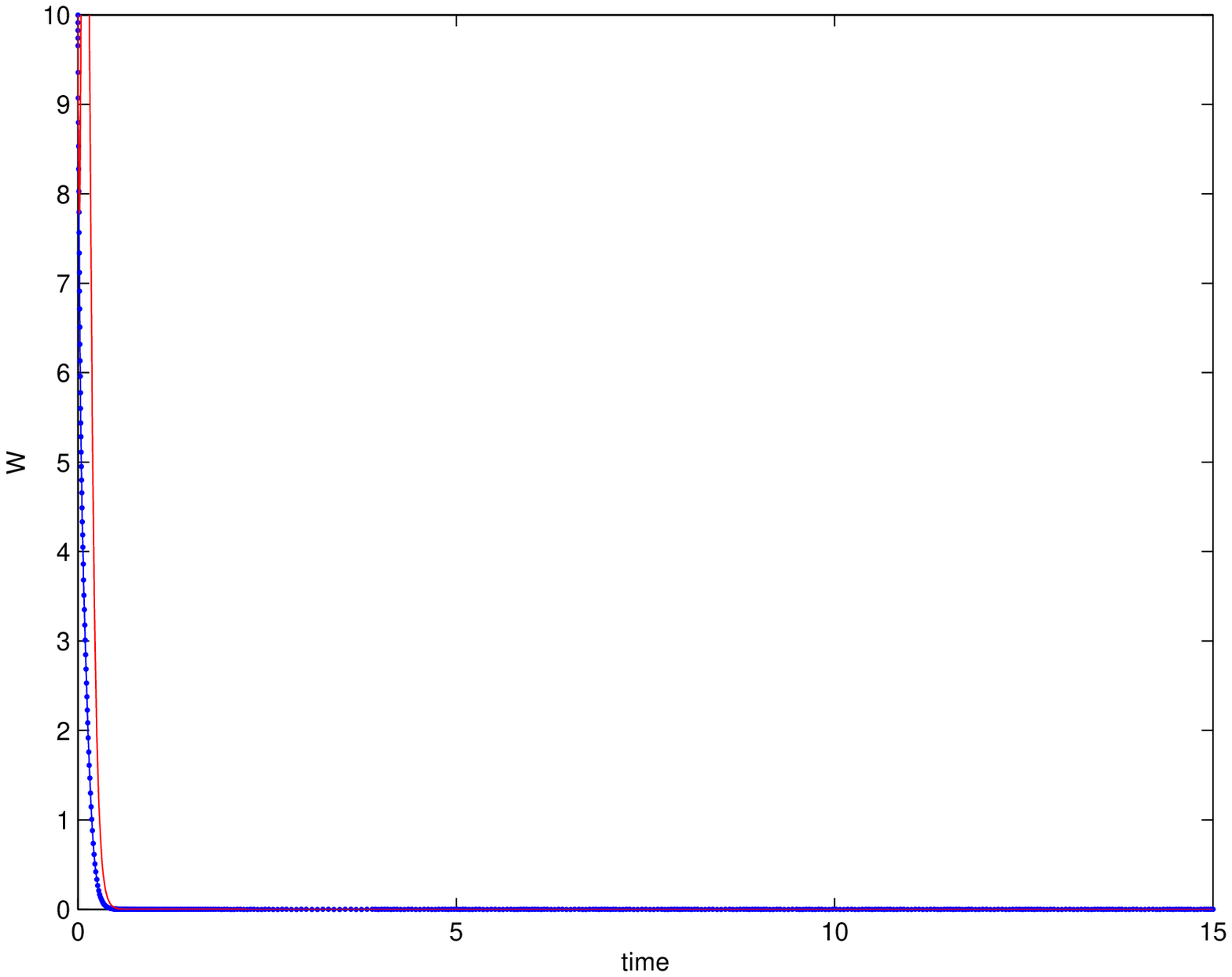}
\end{minipage}%
\hspace{5mm}%
\begin{minipage}[c]{.4\textwidth}
\centering
\includegraphics[width=.78\textwidth]{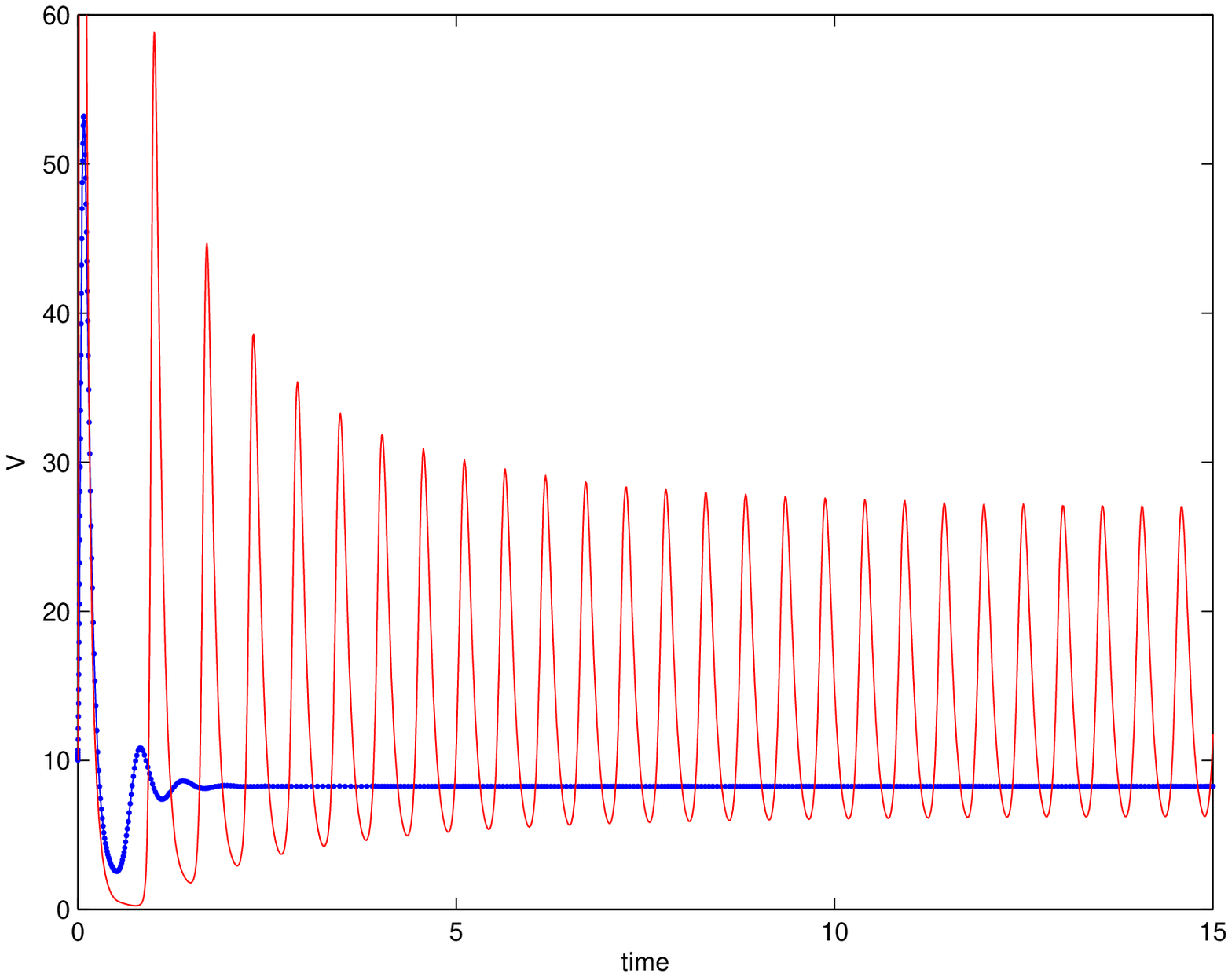}
\end{minipage}
\begin{minipage}[c]{.4\textwidth}
\centering
\includegraphics[width=.78\textwidth]{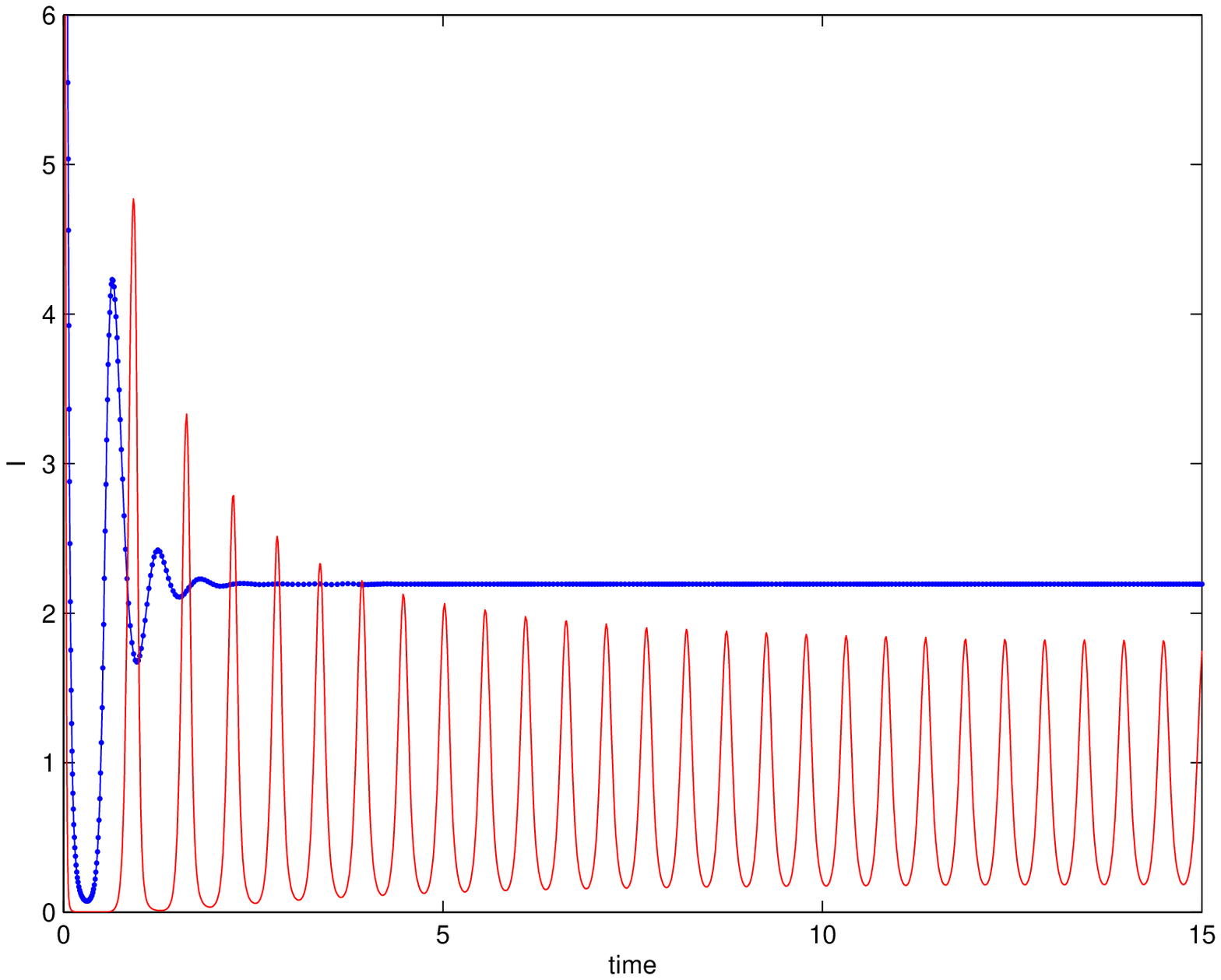}
\end{minipage}
\hspace{5mm}
\begin{minipage}[c]{.4\textwidth}
\centering
\includegraphics[width=.78\textwidth]{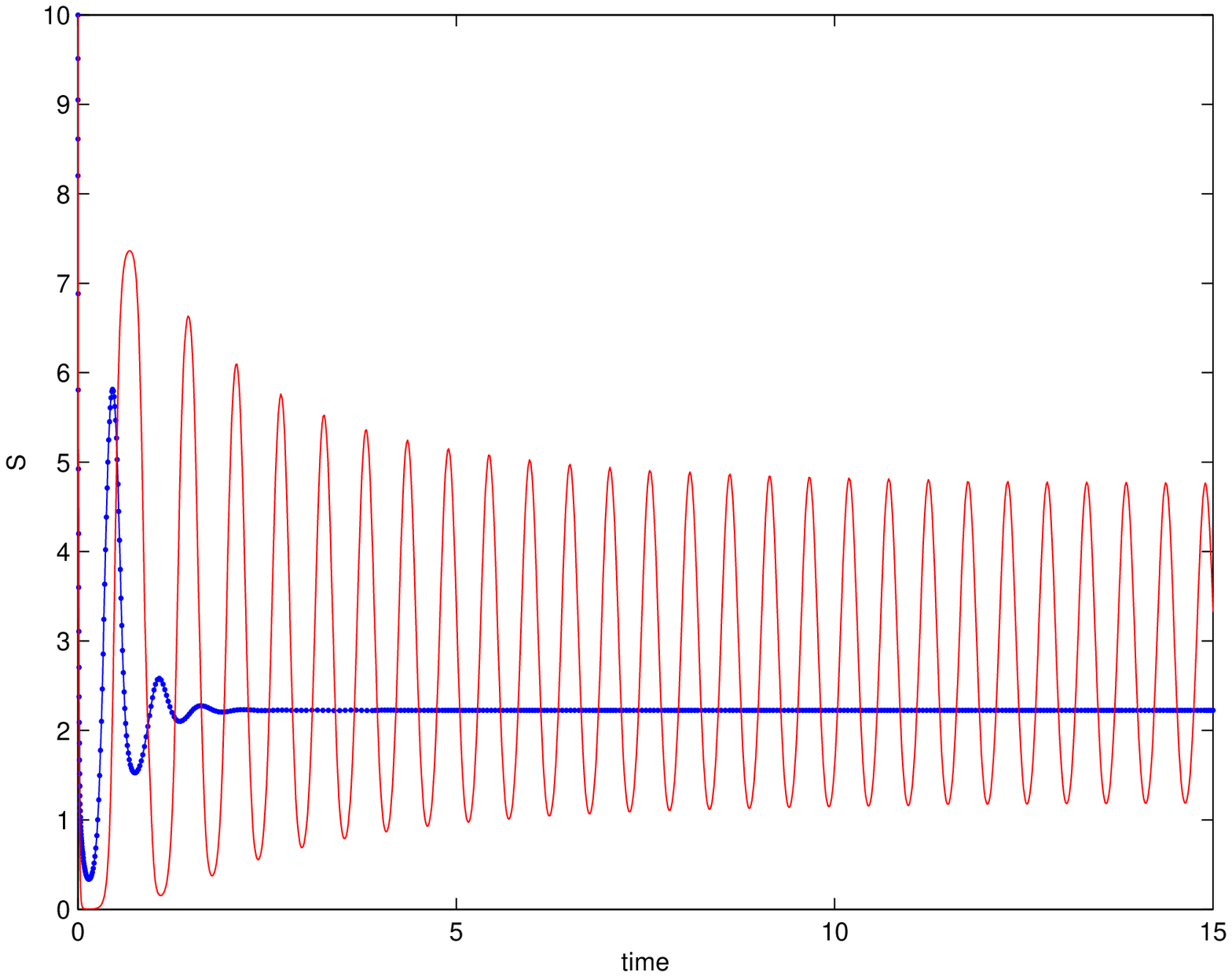}
\end{minipage}
\caption{The plots are obtained for the same parameter set as in Figure \ref{fig_a}, but for $q$:
$q=4$ gives the stable equilibrium (blue line), $q = 11$ provides the limit cycles (red line).}
\label{fig_b}
\end{figure}

We have studied also the system's behavior at this equilibrium point in terms of the hunting rate of the intermediate
predator on the infected prey. The results show that there is a Hopf bifurcation for which limit cycles arise,
Figure \ref{fig_b}. The parameters are the same as in Figure \ref{fig_a}, but for the parameter $q$, which is chosen
as $q=4$ as in the former Figure, as well as $q=11$, the value for which the persistent oscillations are shown.

Endemic coexistence of all the trophic levels is given by
the equilibrium $\mathbf{E}^* \equiv (W^*,V^*,I^*,S^*)$, the population levels of which can be
explicitly evaluated, letting $Z=n m a+a \gamma p+a p \nu+K \beta n m+K \beta p \nu>0$, as
\begin{eqnarray*}
V^* &=& \frac{m}{p},\quad S^* = \frac{n m+\gamma p+\nu p}{\beta p},\\
I^* &= &\frac1{\beta p Z} \left[{-a n^2 m^2-2 a n m \gamma p-2 a n m \nu p-a \gamma^2 p^2-2 a \gamma p^2 \nu-a \nu^2 p^2} \right. \\
& & \left. {-c m^2 K \beta n-c m K \beta \gamma p-c m K \beta \nu p+a K \beta p n m+a K \beta p^2 \gamma+a K \beta p^2 \nu}\right] \\
W^* 
&=& \frac1{h\beta p Z} \left[ {a(e-q)(mn + \gamma p + \nu p)^2+ \beta(qK(ap-cm) - pla)(mn + \gamma p + \nu p) +} \right. \\
&& \left.{+ p \beta K (\gamma e - \beta l) (mn+p \nu) + e \beta K (mn + \nu p)^2} \right]
\end{eqnarray*}
Note indeed that since $W\ne 0$, from the first equilibrium equation $V^*$ is obtained explicitly, and then
from the third equilibrium equation we get explicitly also the value of $S^*$.
In view of the fact that $V^*>0$ and $S^*>0$,
feasibility easily holds if we require 
\begin{equation}\label{E*_feas}
e\ge q, \quad qKap \ge qKcm + pla, \quad \gamma e \ge \beta l.
\end{equation}

\begin{figure}[htbp]
\centering
\title{Coexistence equilibrium}\\
\begin{minipage}[c]{.40\textwidth}
\centering
\includegraphics[width=1\textwidth]{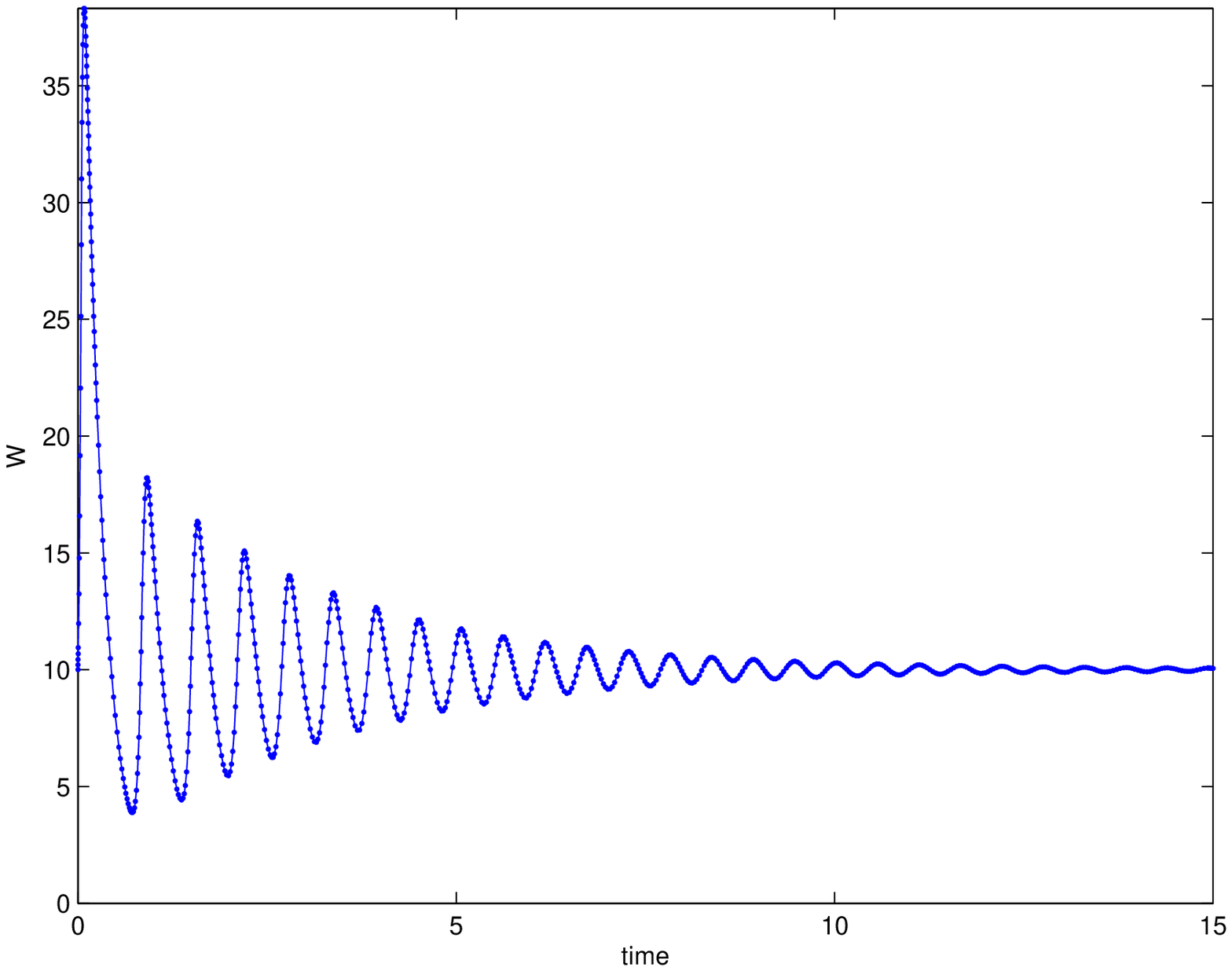}
\end{minipage}
\hspace{10mm}
\begin{minipage}[c]{.40\textwidth}
\centering
\includegraphics[width=1\textwidth]{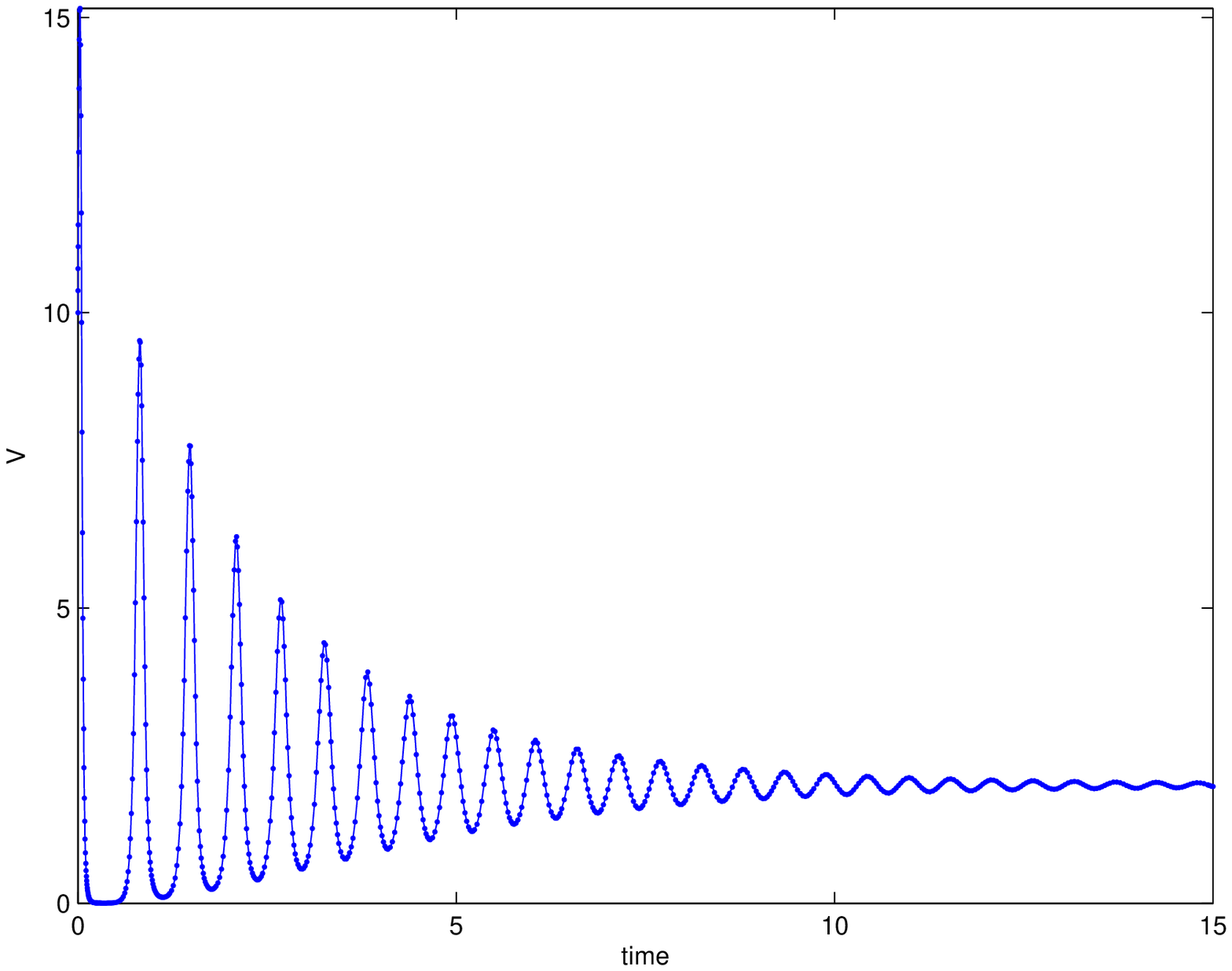}
\end{minipage}
\begin{minipage}[c]{.40\textwidth}
\centering
\includegraphics[width=1\textwidth]{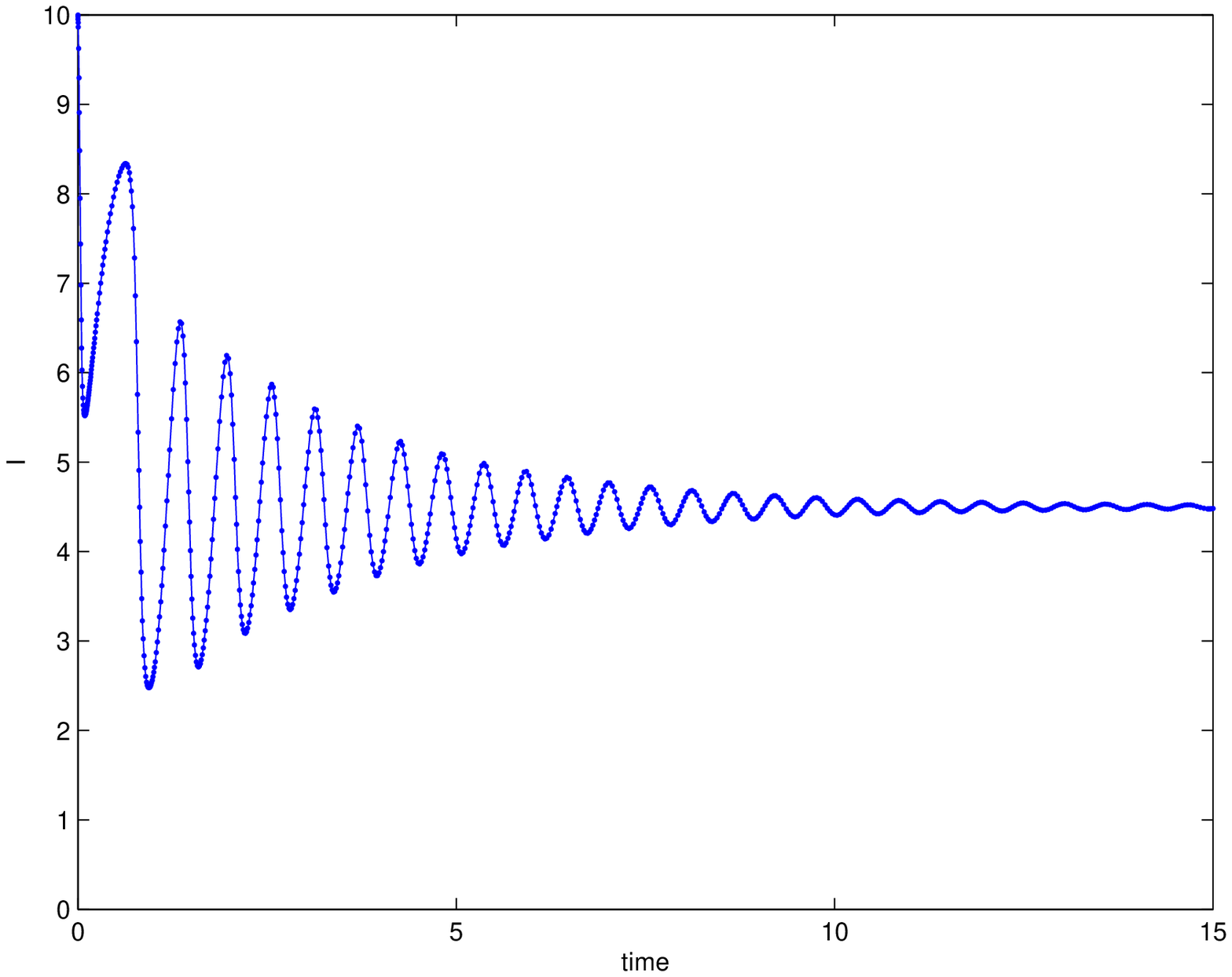}
\end{minipage}
\hspace{10mm}
\begin{minipage}[c]{.40\textwidth}
\centering
\includegraphics[width=1\textwidth]{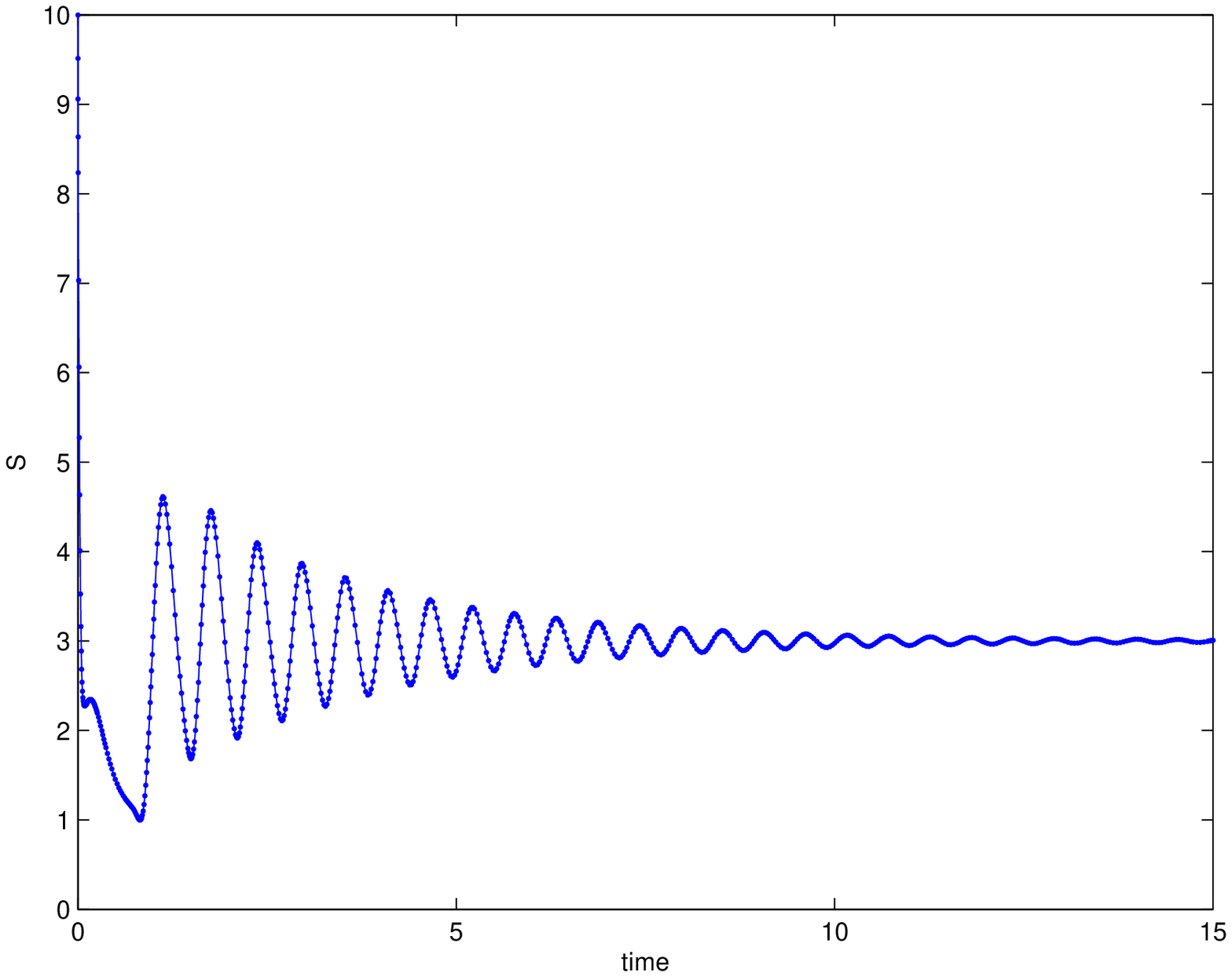}
\end{minipage}
\caption{Coexistence achieved for the parameter values
$a = 20$, $K = 10$, $m = 4$, $p = 2$, $l = 1$, $e = 1$, $h = 2$, $q = 4$, $\beta = 1$, $n = 1$, $c = 1$, $\gamma = 1$, $\nu=0$.}
\label{fig_c}
\end{figure}

This equilibrium can be achieved stably, as shown in Figure \ref{fig_c} for the hypothetical parameter set
$a = 20$, $K = 10$, $m = 4$, $p = 2$, $l = 1$, $e = 1$, $h = 2$, $q = 4$, $\beta = 1$, $n = 1$, $c = 1$, $\gamma = 1$, $\nu=0$.

\begin{figure}[htbp]
\centering
\title{Persistent oscillations of all the populations
}\\
\includegraphics[width=10cm]{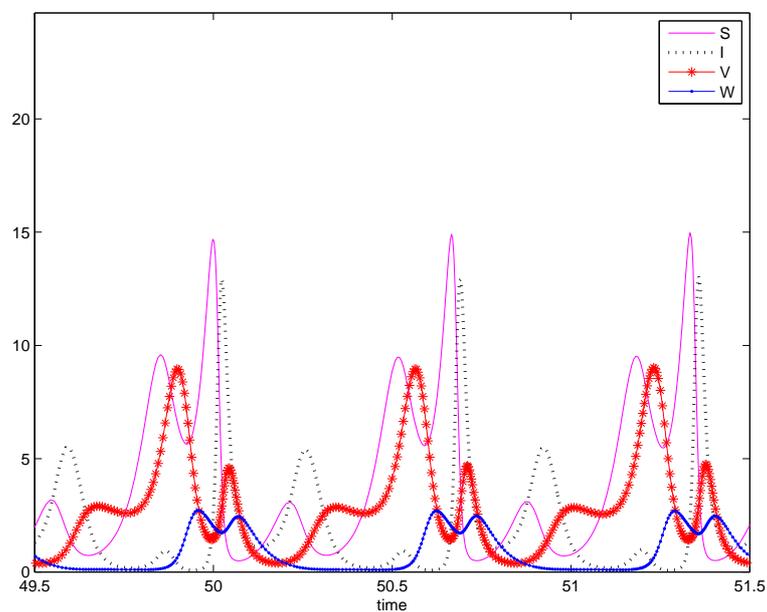}
\caption{The plots are obtained for the parameter values
$m = 20$, $p = 7$, $l = 5$, $e = 2$,
$h = 15$, $q = 5$, $\beta = 9$,
$n = 12$, $\gamma = 0.2$, $\nu = 1.5$,
$a = 30$, $K = 220$, $c = 4$.}
\label{fig_d_new}
\end{figure}

A different choice of the parameter set,
$m = 20$, $p = 7$, $l = 5$, $e = 2$,
$h = 15$, $q = 5$, $\beta = 9$,
$n = 12$, $\gamma = .2$, $\nu = 1.5$,
$a = 30$, $K = 220$, $c = 4$,
leads instead to persistent oscillations in all the ecosystem's populations, Figure \ref{fig_d_new}.
Oscillations are observed also for these parameters
$a = 30$, $K = 120$, $m = 20$, $p = 7$, $l = 1.9$,
$e = 2$, $h = 10$, $q = 5$, $\beta = 9$, $n = 3$, $c = 4$, $\gamma = 0.2$, $\nu=1.5$,
figure not shown.

{\textbf{Remark 4.}} The question of global stability for ecoepidemic food chains deserves
further investigations, as it is not
our main issue in this paper. We just remark that a straightforward application of
the technique of \citet{HS} p. 63 for $\mathbf{E}^*$
here does not work. Even in the easier case in which $\gamma =0$,
it leads to the parameter relationship $cq(a+K \beta)=\beta eKn$, which in general
cannot be satisfied.

\subsection{Bistability}

We now show first that some equilibria combinations cannot possibly stably sussist together.

In view of Remark 2, 
$\mathbf{E}_2$ and $\mathbf{E}_4$ cannot both be feasible and stable
for a given parameter set in view of the stability condition (\ref{E2_stab}) for the former and the feasibility of the
latter (\ref{E4_feas}), which contradict each other.
The same situation occurs also for $\mathbf{E}_2$ and $\mathbf{E}_5$, since once again stability of the former (\ref{E2_stab}) conflicts with feasibility
of the latter, see condition (\ref{E5_feas}), as mentioned in Remark 3.

Again similarly the same happens for
$\mathbf{E}_2$ and $\mathbf{E}_3$,
since feasibility of the latter (\ref{E3_feas}) contains $ap>mc$ and entails 
\begin{equation}\label{E3_feas_bis}
K > K \left( 1 -\frac{mc}{ap}\right) > \frac{l}{e},
\end{equation}
which contradicts (\ref{E2_stab}).

We find the same feature once more for $\mathbf{E}_4$ and $\mathbf{E}_3$.
This is stated in Remark 1,
but it can be better seen from the rephrased feasibility of the latter, (\ref{E3_feas_bis}), which is the opposite of
the first stability condition of the former, (\ref{E4_stab}), rephrased as
\begin{equation}\label{E4_stab_bis}
K \left( 1-\frac{mc}{ap}\right) < \frac{l}{e}.
\end{equation}

Instead, the equilibria $\mathbf{E}_4$ and $\mathbf{E}_5$ can both be feasible and stable.
Indeed, feasibility and stability of $\mathbf{E}_4$, i.e. (\ref{E4_feas}) and (\ref{E4_stab}), require
\begin{equation}\label{eq:p4p5coesistenzap4}
\frac{l}{e} < \frac{na+\gamma c+ c \nu}{\beta c K + a n}K,\quad K \left( 1- \frac{mc}{pa}\right) <  \frac{l}{e} \le K,
\end{equation}
while the corresponding conditions (\ref{E5_feas}) and (\ref{E5_stab}) for $\mathbf{E}_5$ entail
\begin{equation}\label{eq:p4p5coesistenzap5}
\frac{\gamma+\nu}{\beta} < K < \frac{a(\gamma+\nu) + K \beta \mu} {aq(\gamma+\nu)} \left[ l-\frac {e(\gamma+\nu)}{\beta} \right]
+ \frac {\gamma+\nu} {\beta} .
\end{equation}

For studying $\mathbf{E}_4$, we now consider the function
$$
f(n) = \frac{na+\gamma c}{\beta c K+ a n} K
$$
which is a hyperbola, increasing toward the horizontal asymptote $y=K$ from $f(0)=\gamma \beta^{-1}$ for $n\ge 0$
when $K > \gamma \beta^{-1}$ and decreasing to it conversely.
We need to find the intervals of the independent variable $n$ for which
$le^{-1} < f(n)$. Recalling that $le^{-1} \le K$ by feasibility (\ref{E4_feas}), for $K < \gamma \beta^{-1}$
the inequality will be satisfied for all $n\ge 0$. In the opposite case, the equality $le^{-1} = f(n^*)$ holds for
\begin{equation*}
n^* =\max \left\{ 0, K\frac ca \frac{\beta l -\gamma e}{eK-l} \right\} \ge 0.
\end{equation*}
and therefore the inequality is satisfied for every $n> n^*$.

As for $\mathbf{E}_5$, we introduce the function
\begin{equation*}
g(q) = \frac{a(\gamma+\nu) + K \beta \mu} {aq(\gamma+\nu)} \left[ l-\frac {e(\gamma+\nu)}{\beta} \right] + \frac {\gamma+\nu} {\beta} 
\end{equation*}
which is also an equilateral hyperbola, with horizontal asymptote $y=(\gamma+\nu) \beta^{-1}$. By the feasibility condition (\ref{E5_feas}),
this asymptote is always lower than $K$.
The vertical asymptote lies on the vertical axis,
\begin{equation*}
\lim_{q\rightarrow 0^+}{g(q)}=+\infty 
\end{equation*}
if $\beta l> e(\gamma + \nu)$. Thus in this case the hyperbola decreases to the horizontal asymptote and therefore meets the level $K$ at
$$
q^*=\frac{a(\gamma+\nu) + K \beta \mu} {aq(\gamma+\nu)} \left[ l-\frac {e(\gamma+\nu)}{\beta} \right] \frac {\beta} {K\beta -\gamma -\nu}.
$$
In such case then, for $q<q^*$ (\ref{eq:p4p5coesistenzap5}) is satisfied. When $\beta l < e(\gamma + \nu)$ the hyperbola is such that
\begin{equation*}
\lim_{q\rightarrow 0^+}{g(q)}=-\infty 
\end{equation*}
so that it raises up toward the horizontal asymptote, but since this is below the level $K$, (\ref{eq:p4p5coesistenzap5}) can never hold.

\begin{figure}[htbp]
\centering
\title{Bistability of $\mathbf{E}_4$ and $\mathbf{E}_5$ when $K = 6$}\\
\begin{minipage}[c]{.40\textwidth}
\centering
\includegraphics[width=1\textwidth]{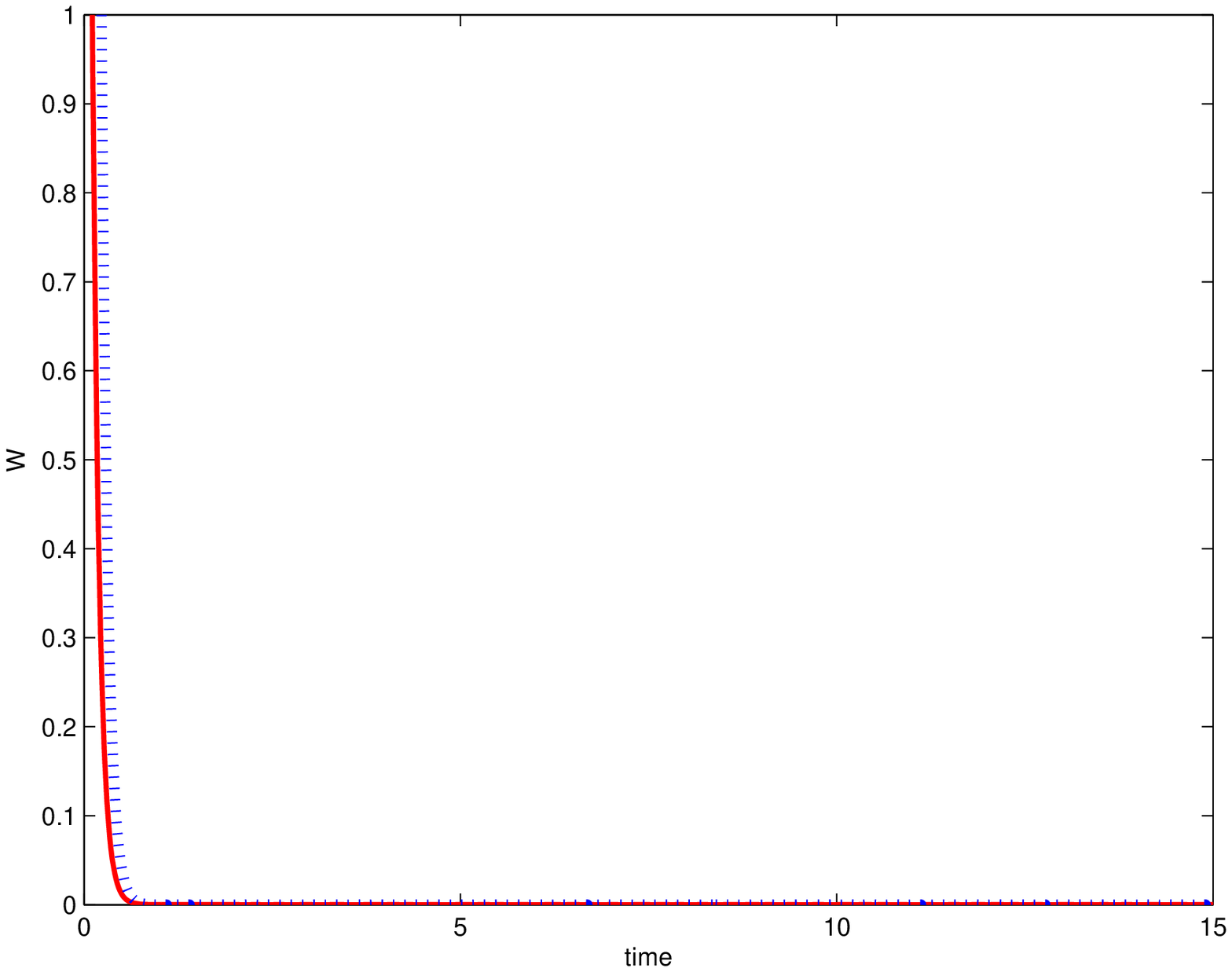}
\end{minipage}
\hspace{10mm}
\begin{minipage}[c]{.40\textwidth}
\centering
\includegraphics[width=1\textwidth]{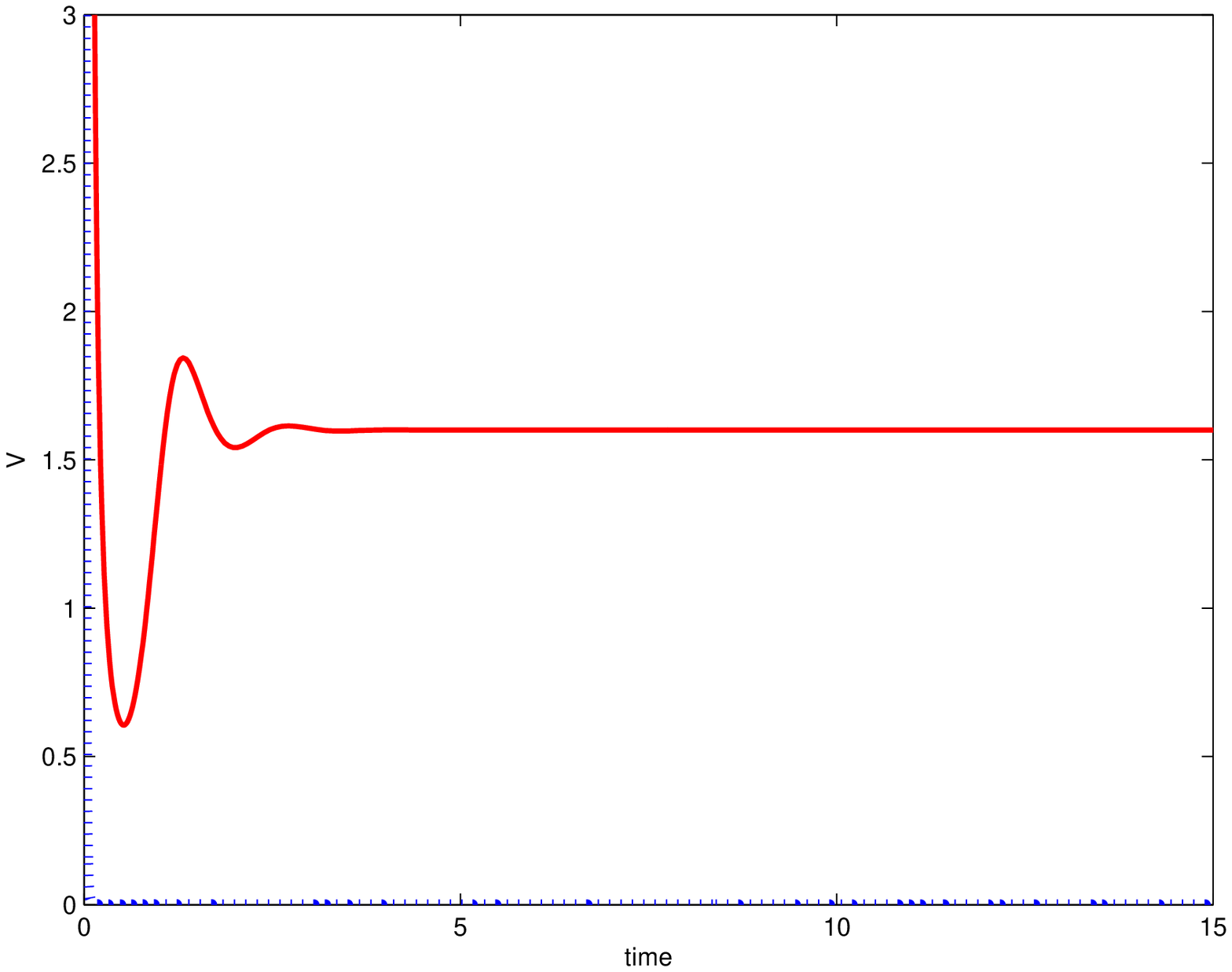}
\end{minipage}
\begin{minipage}[c]{.40\textwidth}
\centering
\includegraphics[width=1\textwidth]{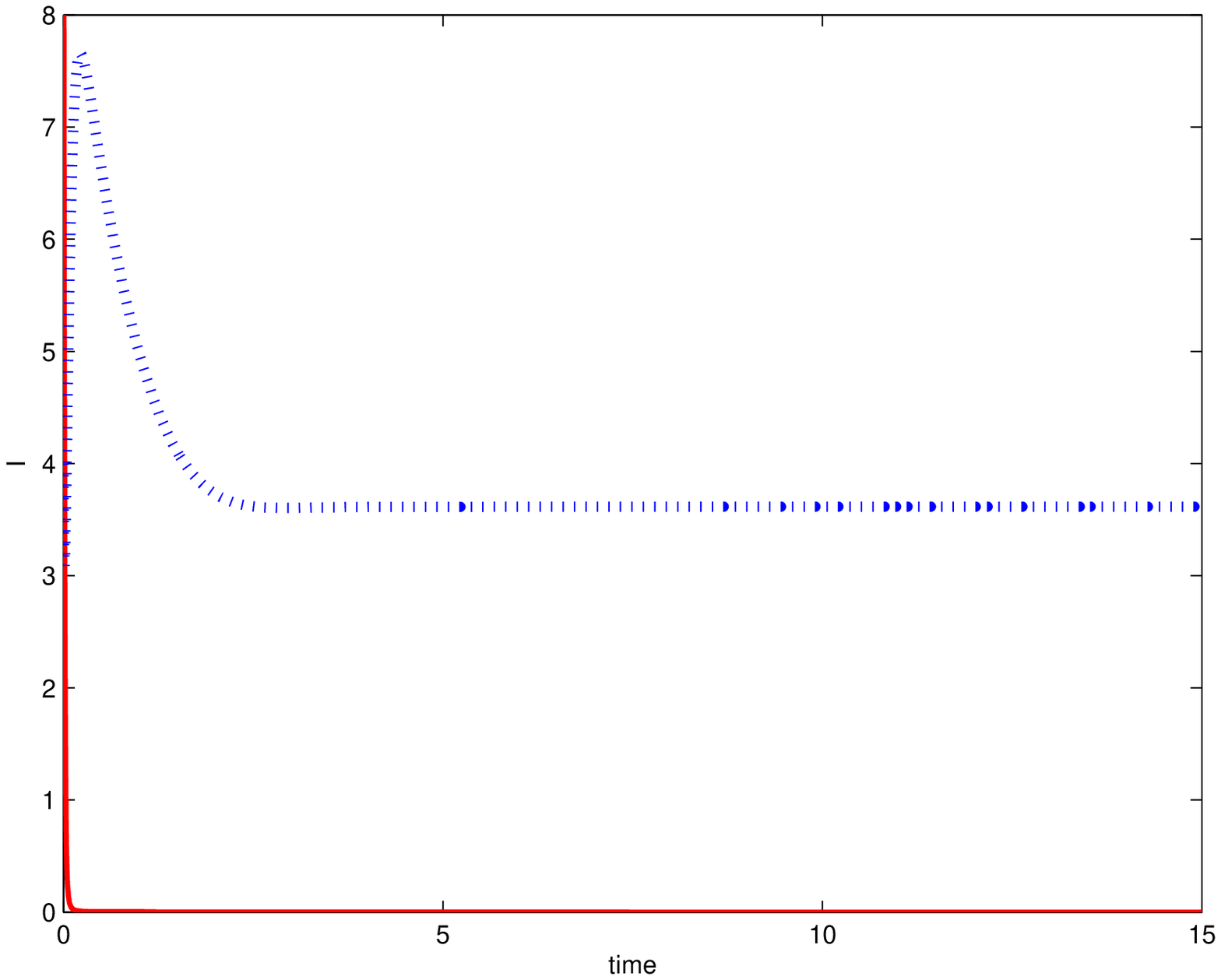}
\end{minipage}
\hspace{10mm}
\begin{minipage}[c]{.40\textwidth}
\centering
\includegraphics[width=1\textwidth]{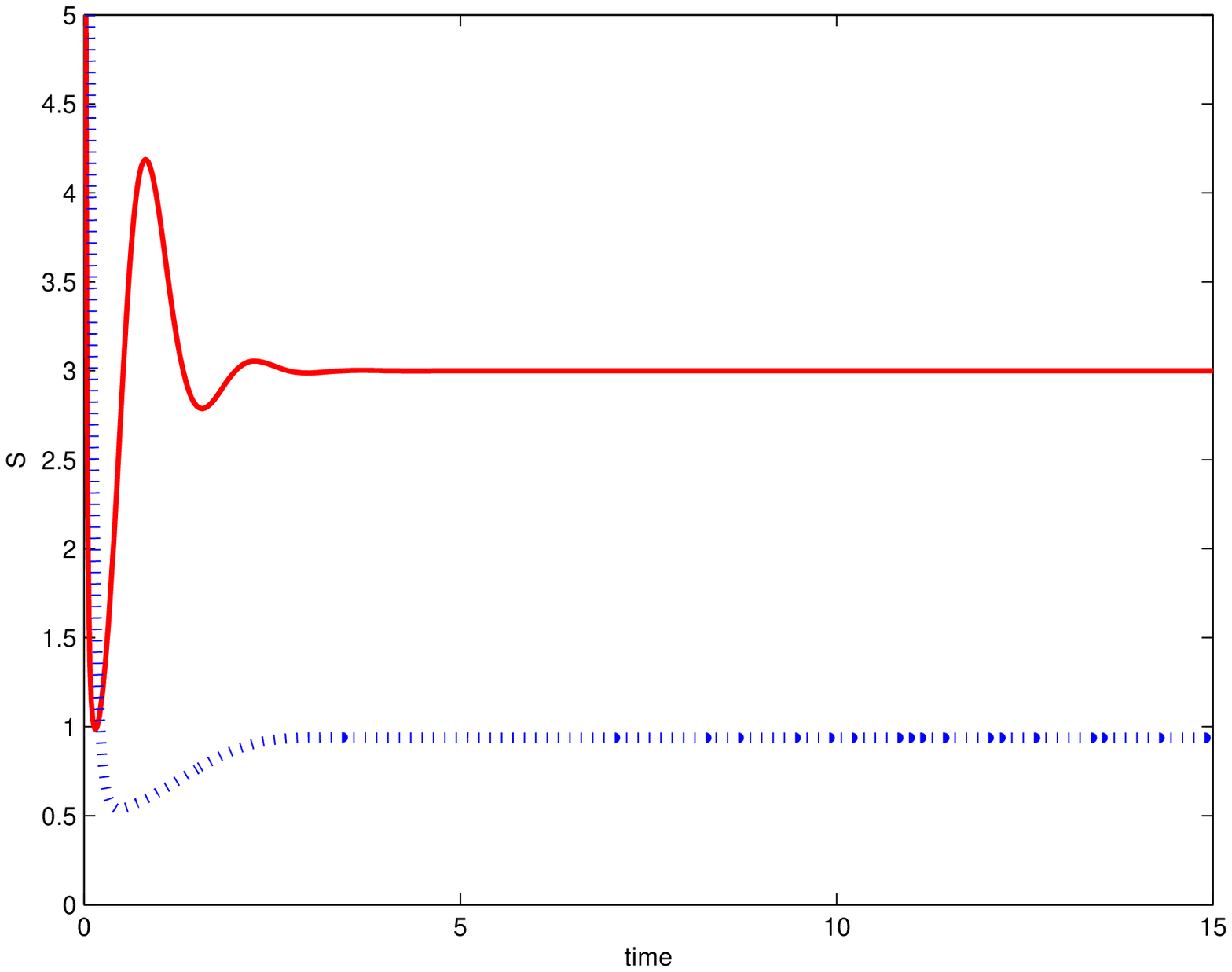}
\end{minipage}
\caption{The other parameters are
$m = 12$, $p = 0.5$, $l = 6$, $e = 2$, $h = 10$, $q = 1$, $\beta = 1.6$, $n = 5$, $\gamma = 1$, $\nu = 0.5$,
$a = 8$, $c = 2.5$.}
\label{fig_j}
\end{figure}

In summary, for coexistence of $\mathbf{E}_4$ and $\mathbf{E}_5$ we need
\begin{equation}\label{coex_4_5}
\frac{l}{e} < \frac{a}{b} < \frac{l}{e} + \frac{mc}{pb},\quad n >n^*,\quad q<q^*,\quad \frac{\gamma}{\beta} < \min \left\{ K, \frac{l}{e} \right\}.
\end{equation}
The bistability is indeed achieved, as can be seen in Figure \ref{fig_j}, for two choices of the initial conditions.

\begin{figure}[htbp]
\centering
\title{Bistability of $\mathbf{E}_5$ and $\mathbf{E}_3$}\\
\begin{minipage}[c]{.40\textwidth}
\centering
\includegraphics[width=1\textwidth]{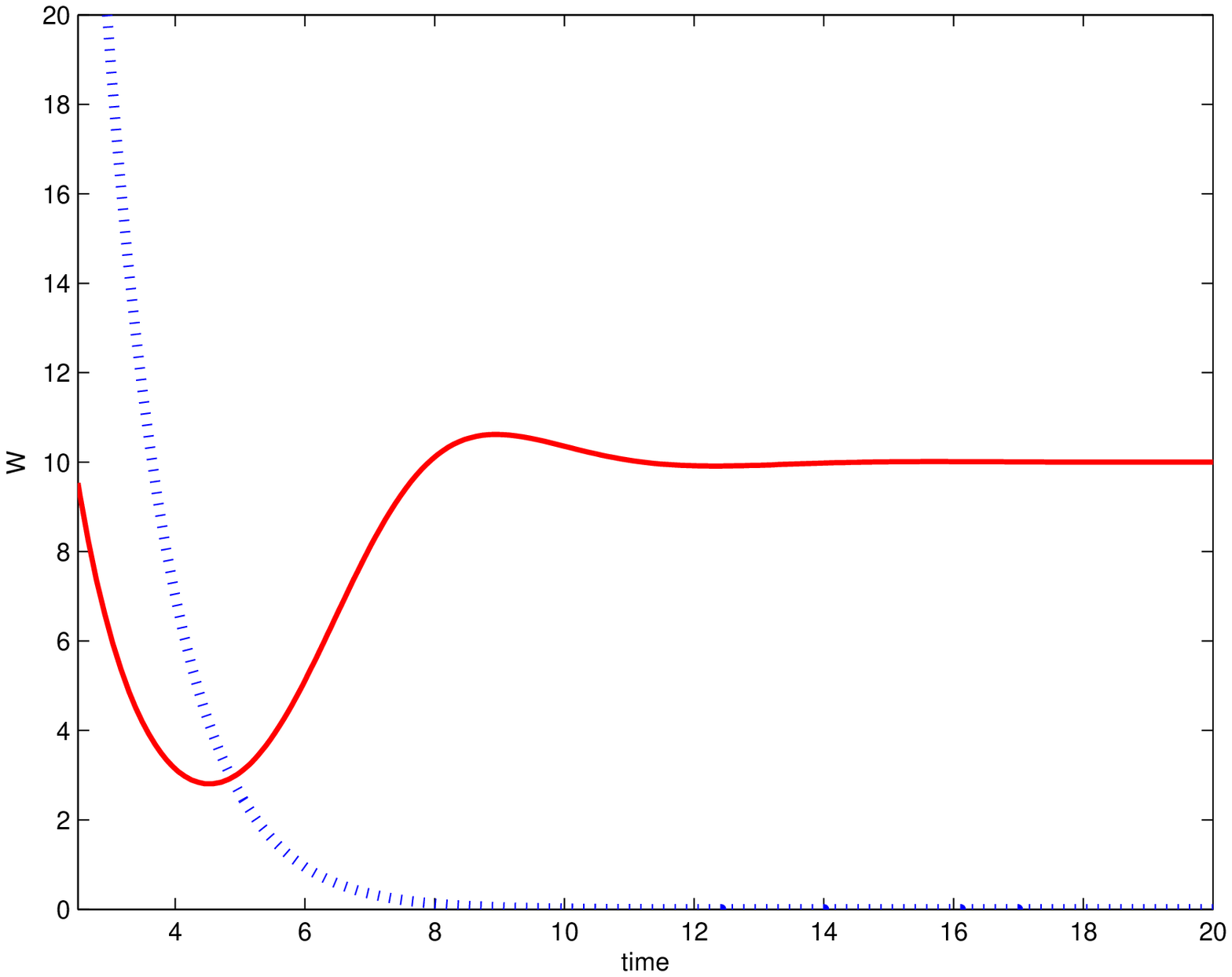}
\end{minipage}
\hspace{10mm}
\begin{minipage}[c]{.40\textwidth}
\centering
\includegraphics[width=1\textwidth]{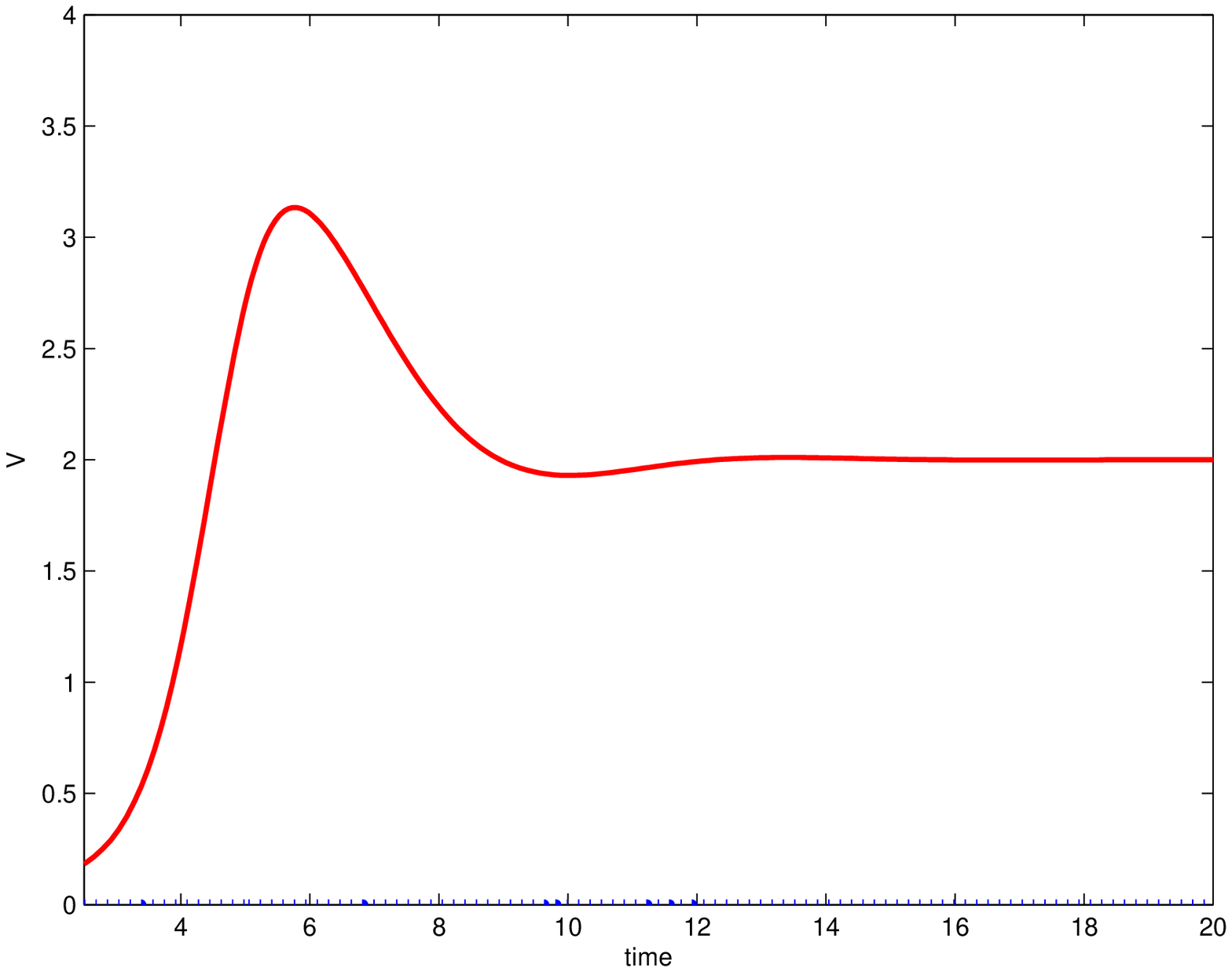}
\end{minipage}
\begin{minipage}[c]{.40\textwidth}
\centering
\includegraphics[width=1\textwidth]{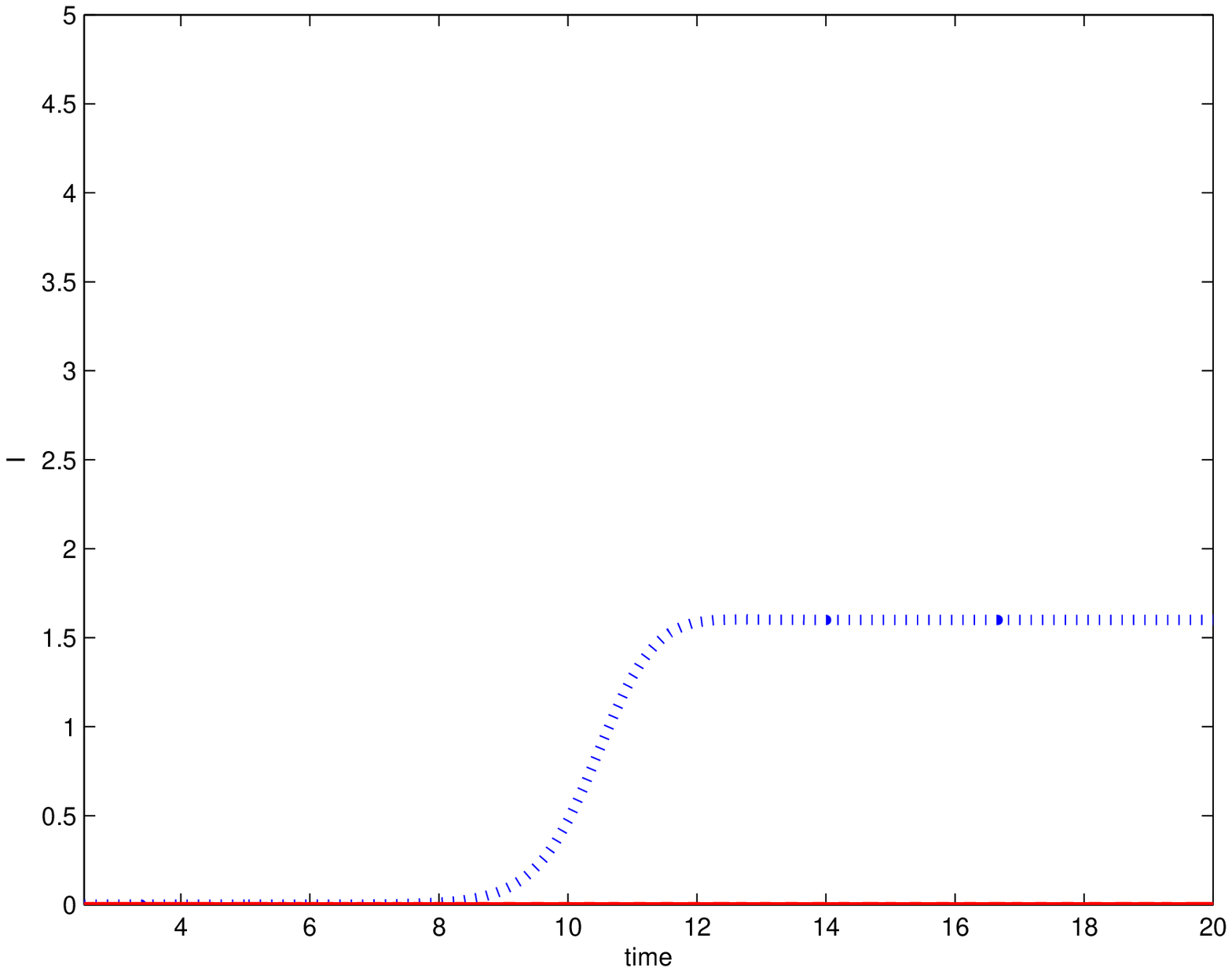}
\end{minipage}
\hspace{10mm}
\begin{minipage}[c]{.40\textwidth}
\centering
\includegraphics[width=1\textwidth]{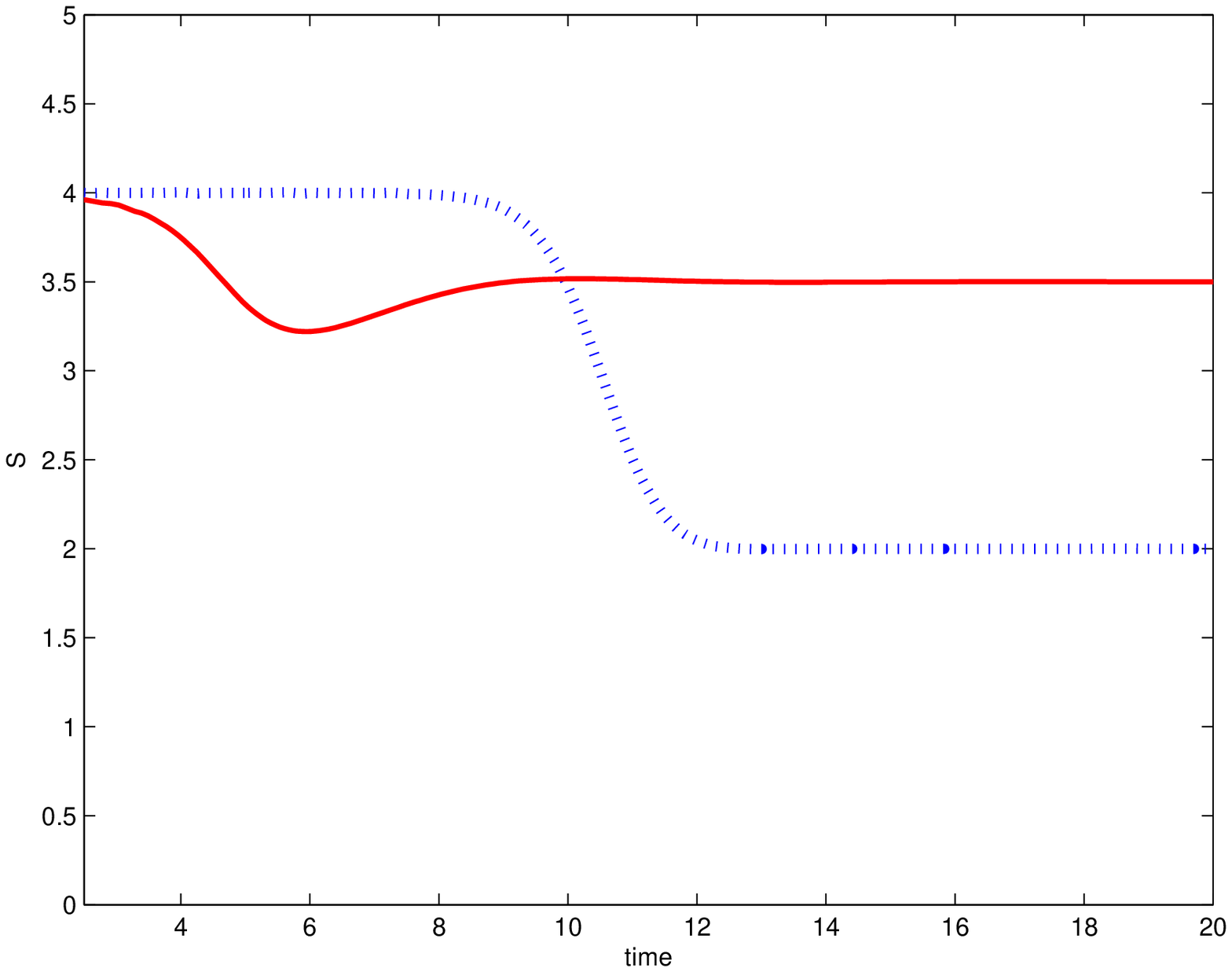}
\end{minipage}
\caption{$\mathbf{E}_5$ (blue) and $\mathbf{E}_3$
(red) are both stable for four different choices of the initial conditions.
Parameter values:
$a = 8$, $K = 4$, $m = 1$, $p = 0.5$, $l = 6$, $e = 2$, $h = 0.1$, $q = 1$, $\beta = 1$, $n = 5$, $c = 0.5$, $\gamma = 1$, $\nu=0$.
}
\label{fig_f}
\end{figure}

Bistability occurs also for the pair of equilibria $\mathbf{E}_5$ and $\mathbf{E}_3$, as shown in Figure \ref{fig_f}
taking four different choices of the initial conditions, for the parameter values
$a = 8$, $K = 4$, $m = 1$, $p = 0.5$, $l = 6$, $e = 2$, $h = 0.1$, $q = 1$, $\beta = 1$, $n = 5$, $c = 0.5$, $\gamma = 1$, $\nu=0$.

\begin{figure}[htbp]
\centering
\title{Bistability of $\mathbf{E}_7$ and $\mathbf{E}^*$}\\
\begin{minipage}[c]{.40\textwidth}
\centering
\includegraphics[width=1\textwidth]{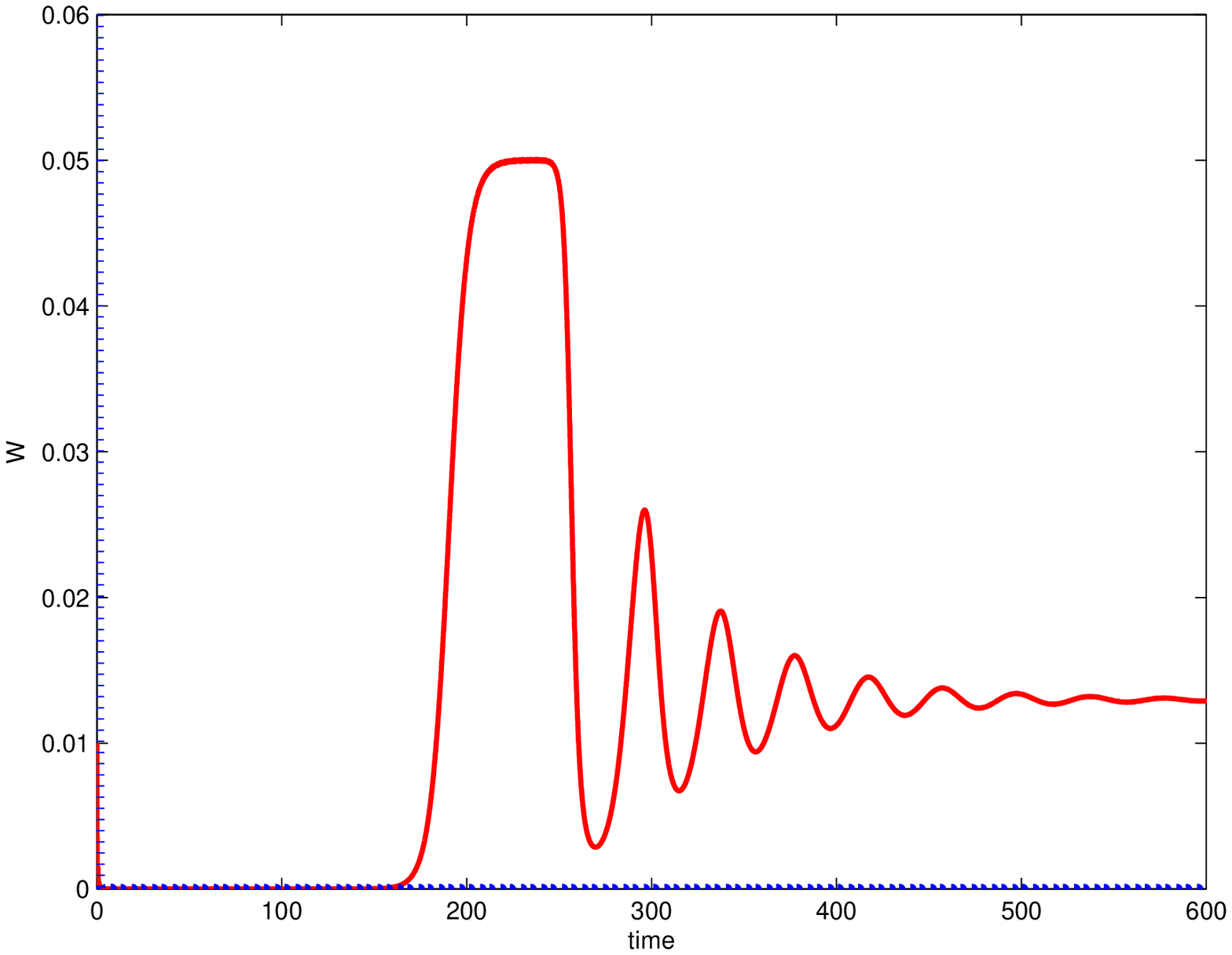}
\end{minipage}%
\hspace{10mm}%
\begin{minipage}[c]{.40\textwidth}
\centering
\includegraphics[width=1\textwidth]{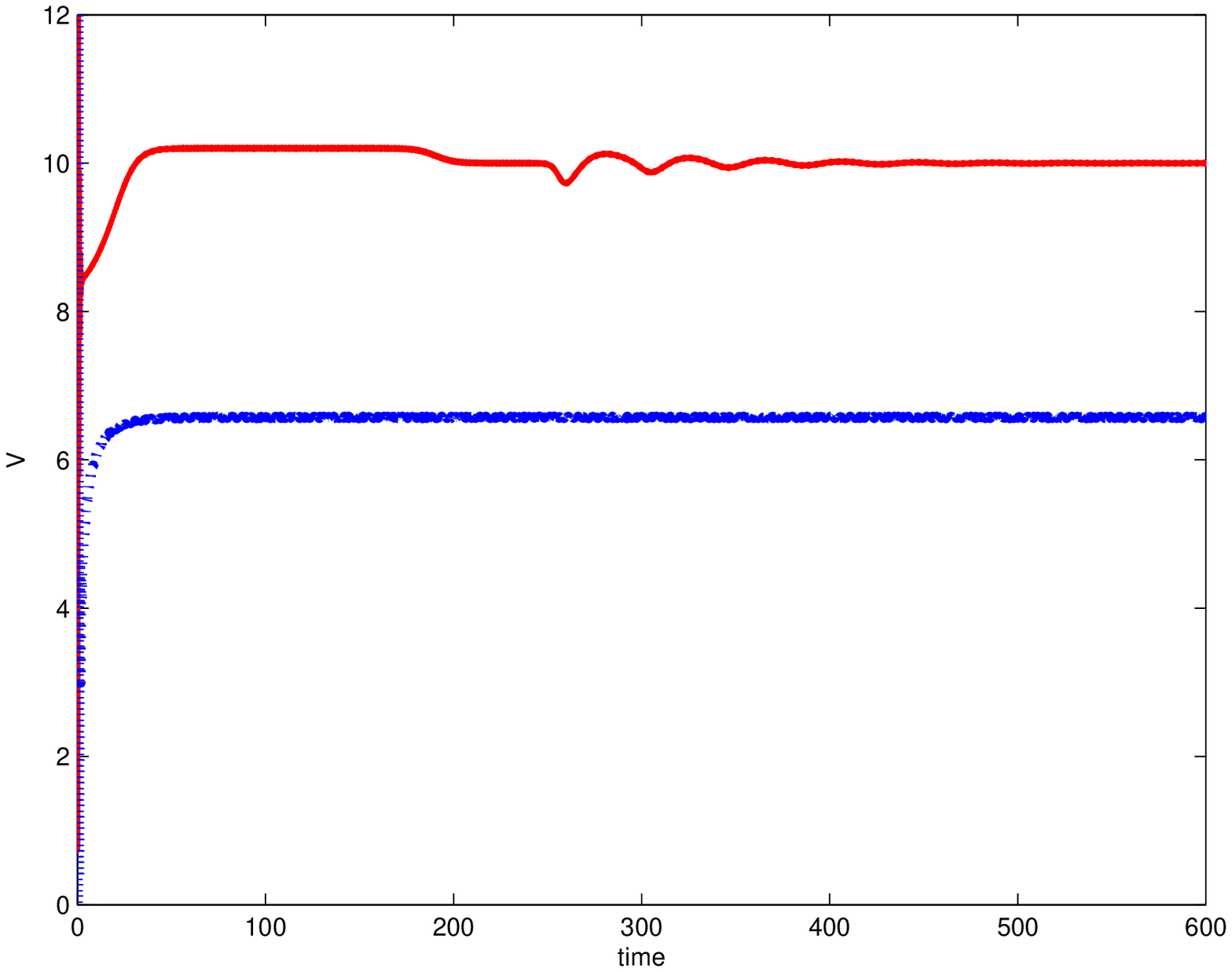}
\end{minipage}
\begin{minipage}[c]{.40\textwidth}
\centering
\includegraphics[width=1\textwidth]{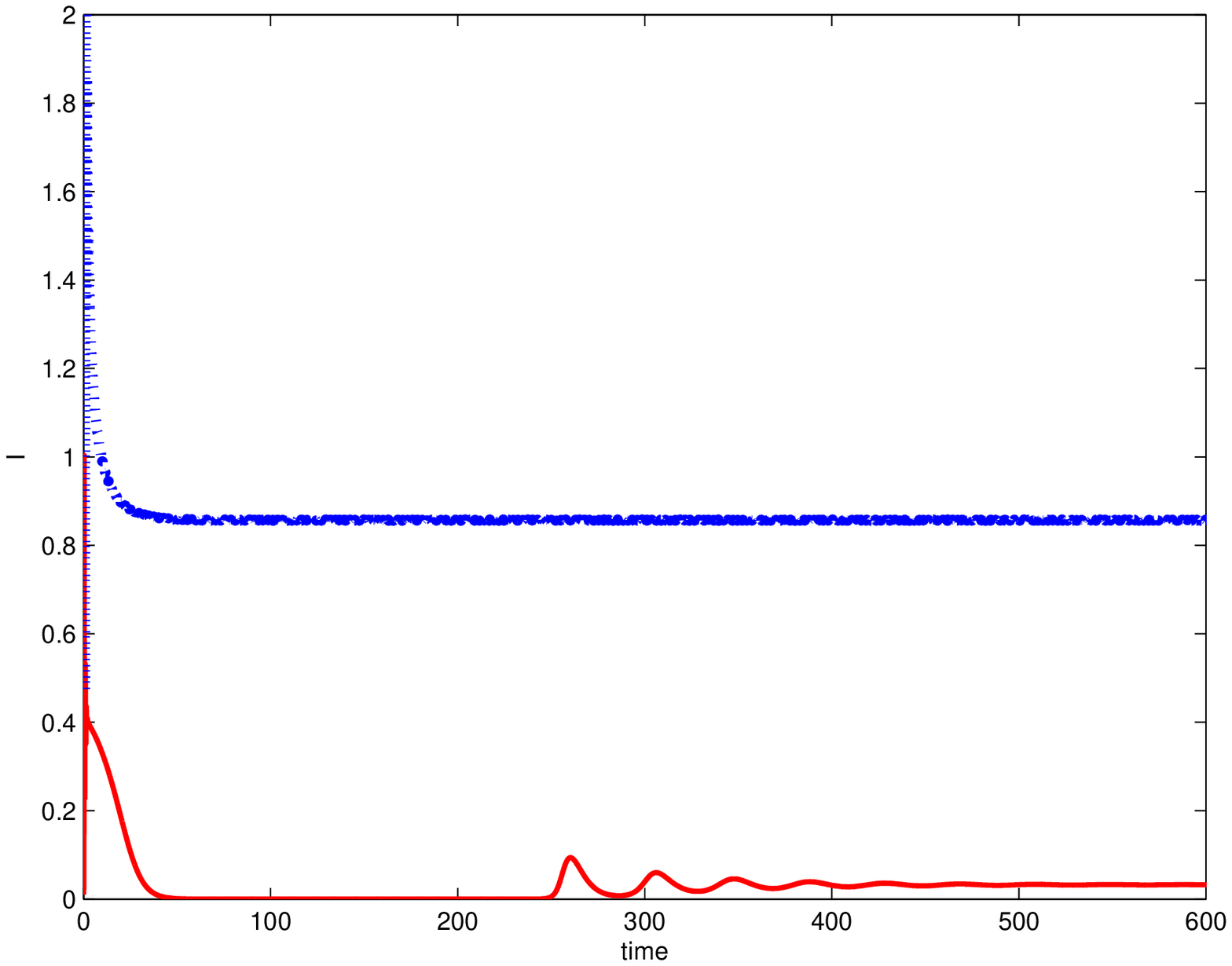}
\end{minipage}%
\hspace{10mm}%
\begin{minipage}[c]{.40\textwidth}
\centering
\includegraphics[width=1\textwidth]{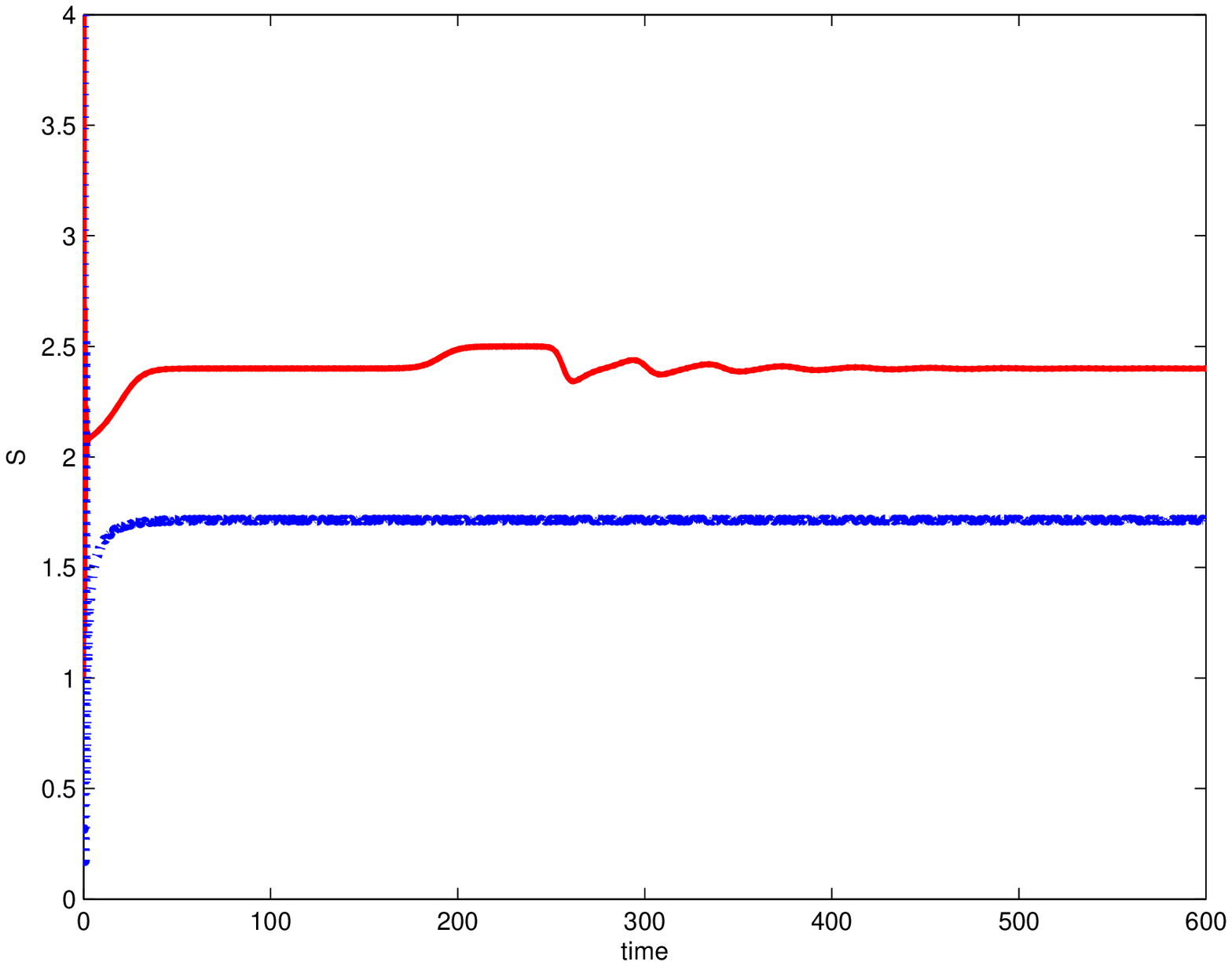}
\end{minipage}
\caption{$\mathbf{E}_7$ (blue) and $\mathbf{E}^*$ (red) are both stable.
Parameters used:
$a = 15$, $K = 7.5$, $m = 10$, $p = 1$, $l = 12$, $e = 5$, $h = 10$, $q = 4$, $\beta = 5$, $n = 1$, $c = 1$, $\gamma = 2$, $\nu=0$.}
\label{fig_g}
\end{figure}

Finally in Figure \ref{fig_g} we show empirically the bistability of the equilibria $\mathbf{E}_7$ and $\mathbf{E}^*$
for the parameters
$a = 15$, $K = 7.5$, $m = 10$, $p = 1$, $l = 12$, $e = 5$, $h = 10$, $q = 4$, $\beta = 5$, $n = 1$, $c = 1$, $\gamma = 2$, $\nu=0$.
 
\begin{figure}[htbp]
\centering
\title{Bistability of $\mathbf{E}_4$ and $\mathbf{E}_7$ for $K = 7$}\\
\begin{minipage}[c]{.40\textwidth}
\centering
\includegraphics[width=1\textwidth]{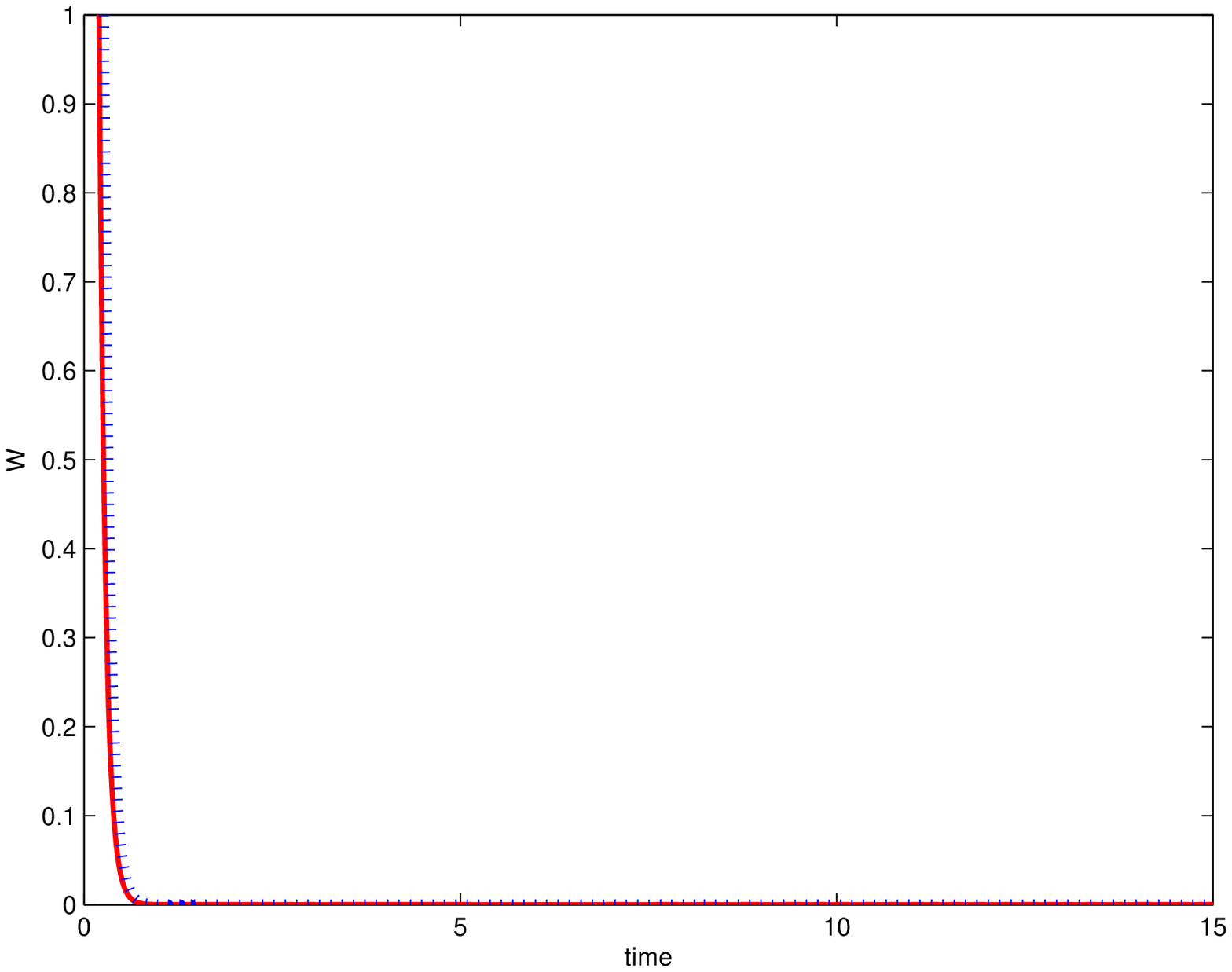}
\end{minipage}
\hspace{10mm}
\begin{minipage}[c]{.40\textwidth}
\centering
\includegraphics[width=1\textwidth]{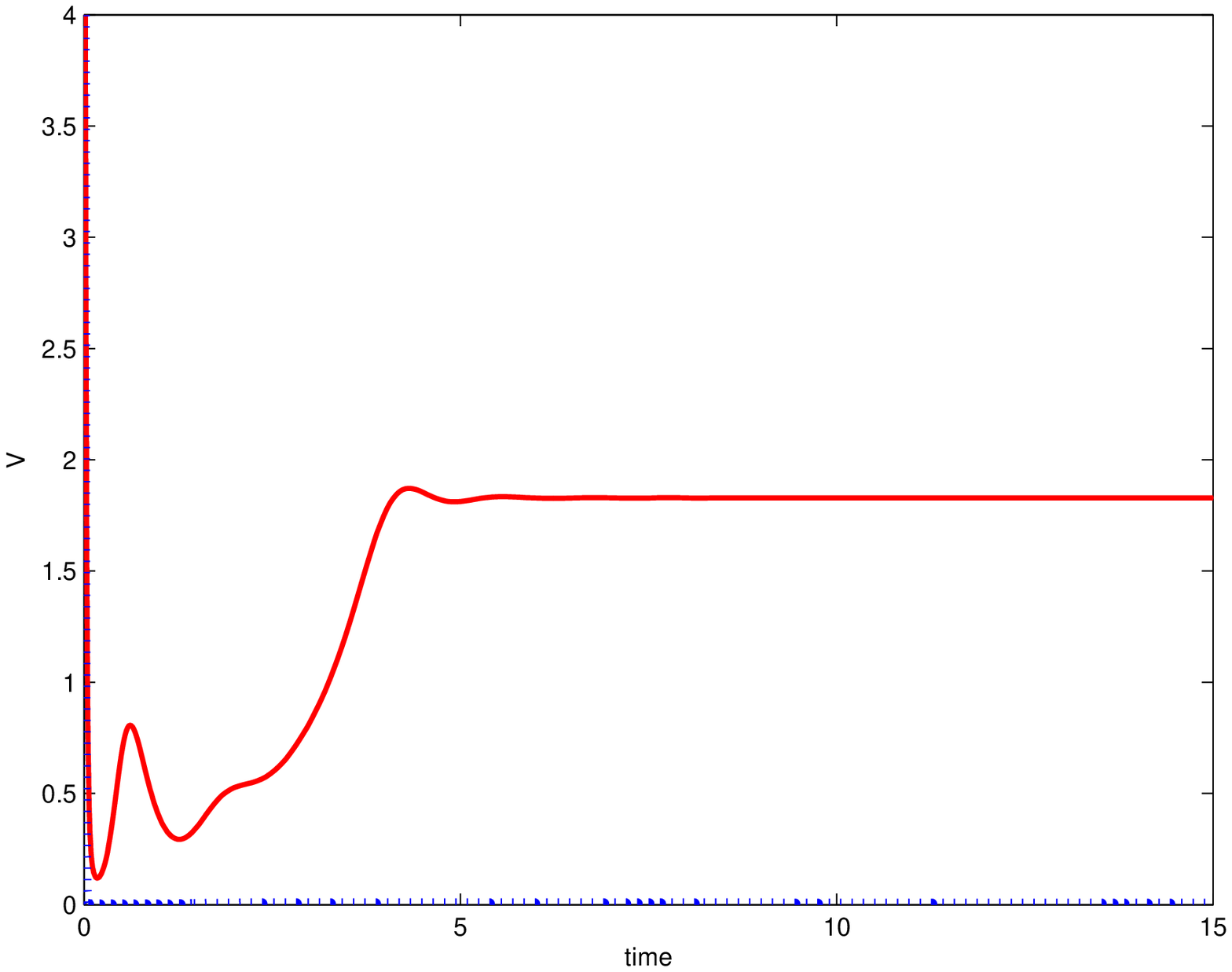}
\end{minipage}
\begin{minipage}[c]{.40\textwidth}
\centering
\includegraphics[width=1\textwidth]{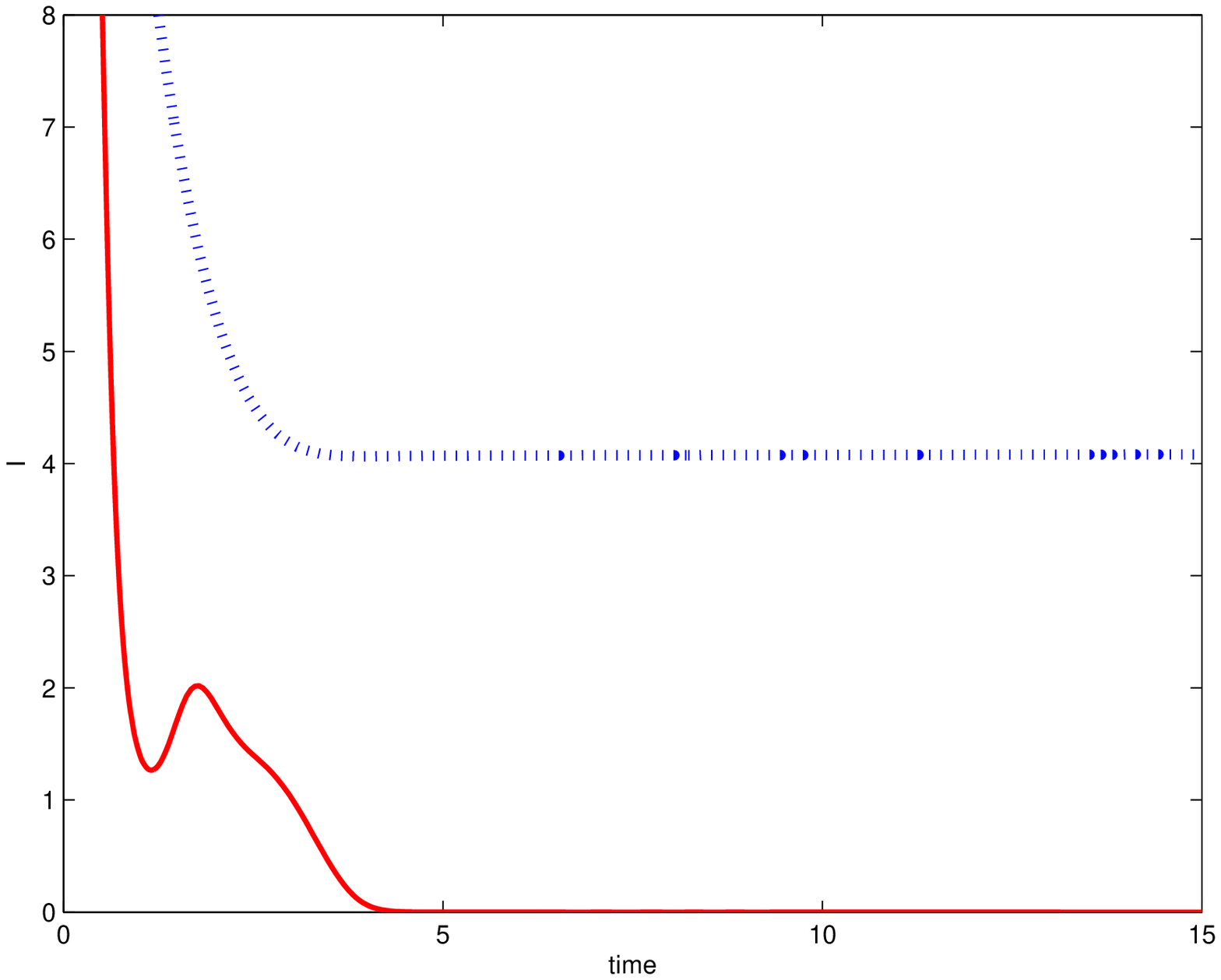}
\end{minipage}
\hspace{10mm}
\begin{minipage}[c]{.40\textwidth}
\centering
\includegraphics[width=1\textwidth]{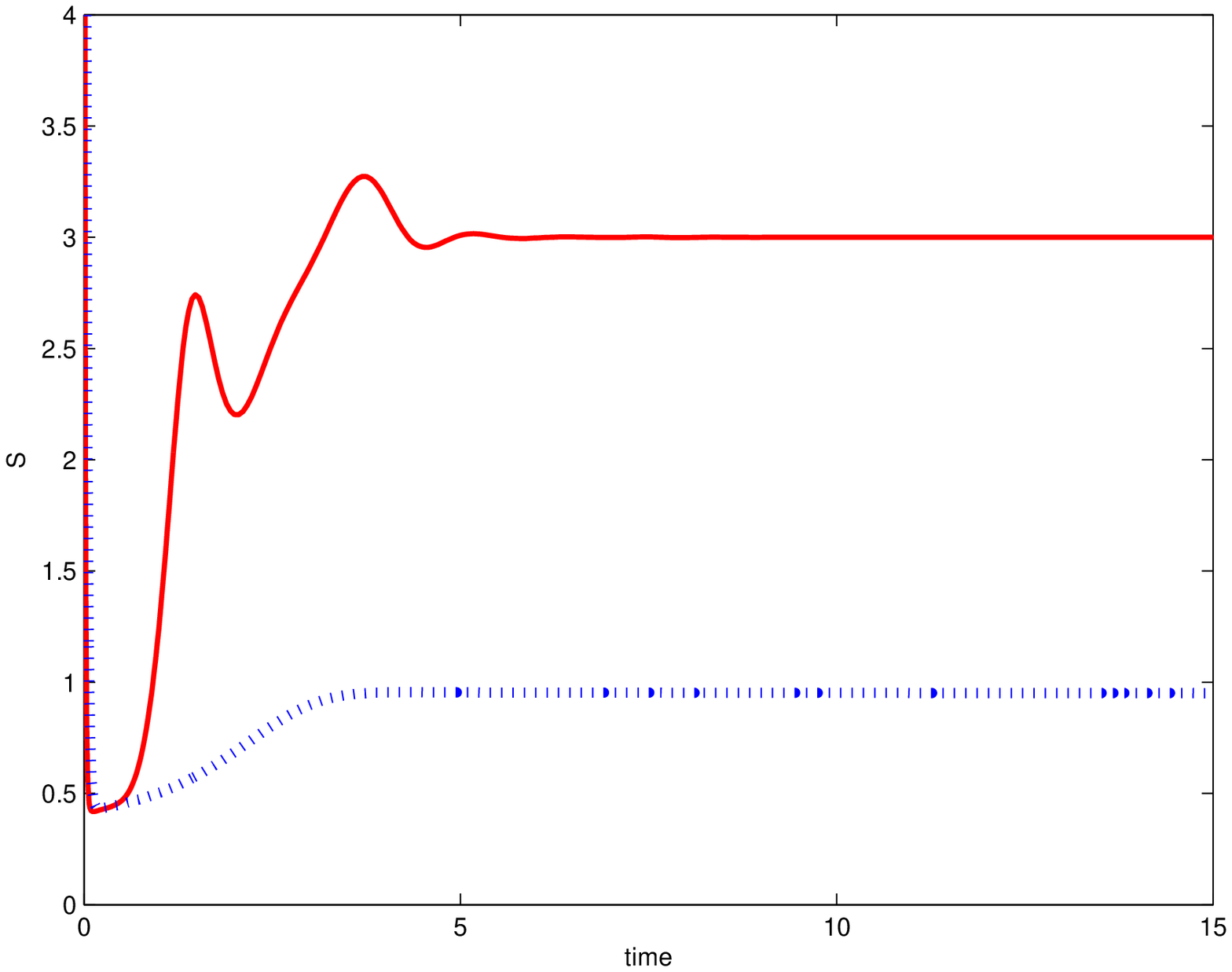}
\end{minipage}
\caption{The other parameters are the same as in Figure \ref{fig_j}.}
\label{fig_h}
\end{figure}

It is interesting also to remark how some of these equilibria behave as $K$ changes.
For $K = 6$, the equilibria $\mathbf{E}_4$ and $\mathbf{E}_5$ coexist, Figure \ref{fig_j}.
The other parameters are chosen as follows:
$m = 12$, $p = 0.5$, $l = 6$, $e = 2$, $h = 10$, $q = 1$, $\beta = 1.6$, $n = 5$, $\gamma = 1$, $\nu = 0.5$, $a = 8$, $c = 2.5$.
If we change the carrying capacity to the value $K = 7$ we discover coexistence of the equilibria $\mathbf{E}_4$ and $\mathbf{E}_7$,
see Figure \ref{fig_h}.

\begin{figure}[htbp]
\centering
\title{Separatrix surface}\\
\includegraphics[width=1\textwidth]{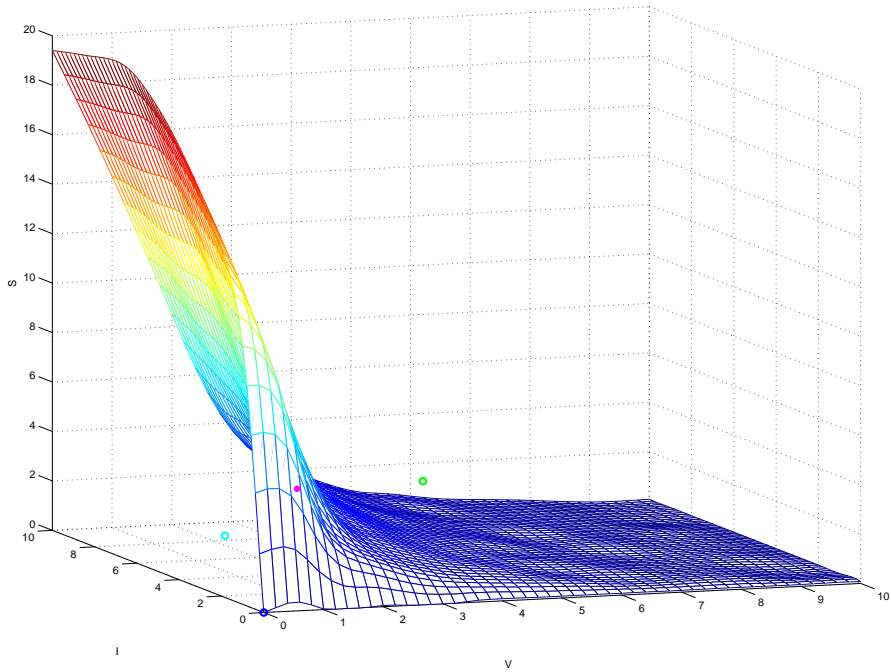}
\caption{The surface separating the basins of attraction of the equilibria $\mathbf{E}_4$,
lying on the coordinate hyperplane $S-V$, and $\mathbf{E}_5$,
lying on the coordinate hyperplane $I-S$,
projected in the phase-subspace $W=0$ for the parameters
$l = 10$, $e = 2$, $q = 1$, $\beta = 1.6$, $n = 5$, $\gamma = 1$, $\nu = 3$, $a = 8$, $K = 6$, $c = 0.5$.
The two equilibria are marked with small green circles on the two sides of the separatrix,
$\mathbf{E}_5$ on the left and $\mathbf{E}_4$ on the right,
the saddle point on the surface by a red circle.}
\label{fig_sep}
\end{figure}

The separatrix for the basins of attraction of the equilibria $\mathbf{E}_4$ and $\mathbf{E}_5$ is
shown in Figure \ref{fig_sep}, in the $W=0$ three-dimensional phase subspace,
for the hypothetical parameter values $l = 10$, $e = 2$, $q = 1$, $\beta = 1.6$, $n = 5$, $\gamma = 1$,
$\nu = 3$, $a = 8$, $K = 6$, $c = 0.5$. The figure is produced using very recently developed
approximation algorithms \citep{ICNAAM,CMMSE,IJCM}.
For further details on the interpolation method, see e.g. \citet{Cav13,Wend05}.

\begin{figure}[htbp]
\centering
\title{Transcritical bifurcation between $\mathbf{E}_3$ and $\mathbf{E}^*$}\\
\begin{minipage}[c]{.40\textwidth}
\centering
\includegraphics[width=1\textwidth]{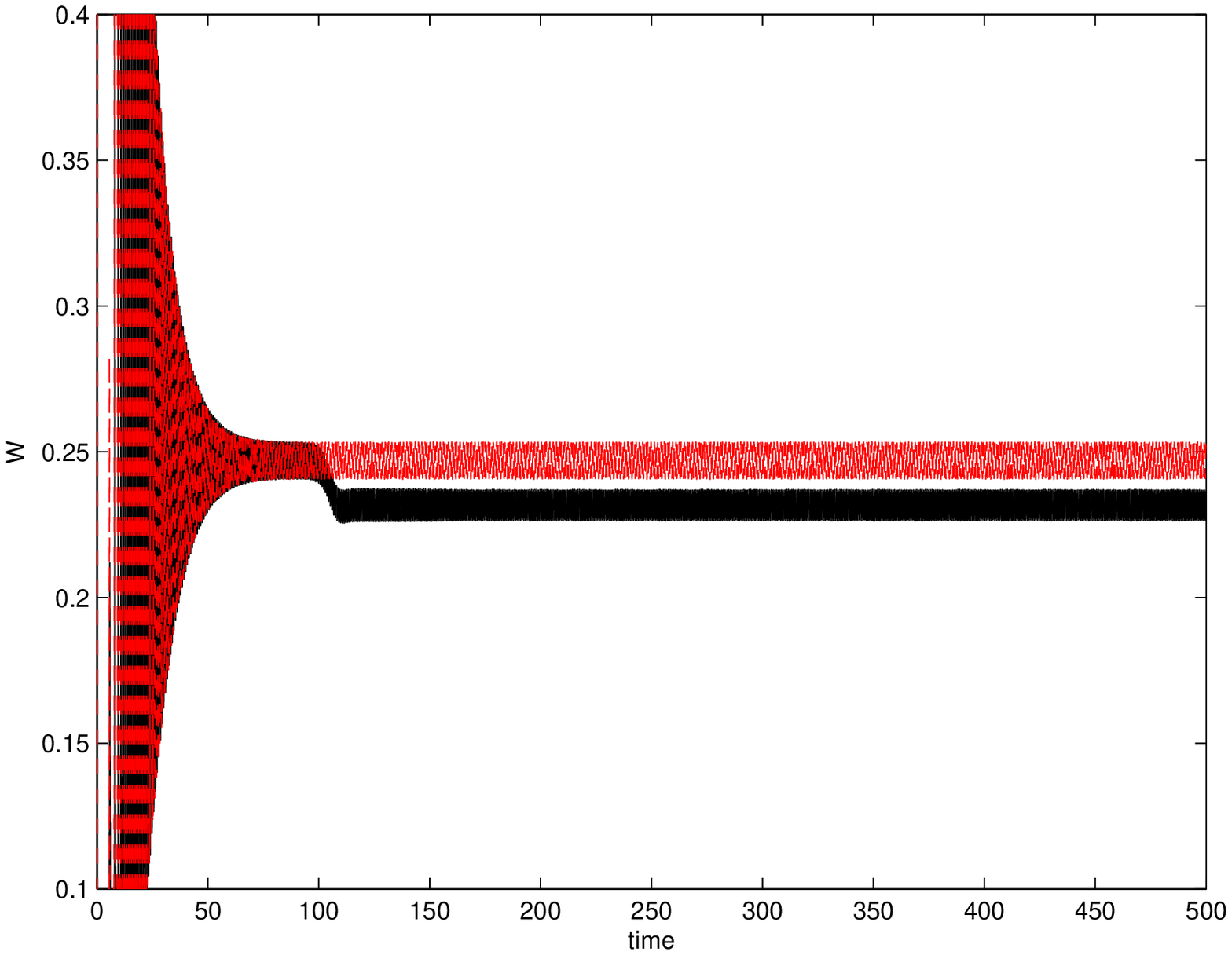}
\end{minipage}
\hspace{10mm}
\begin{minipage}[c]{.40\textwidth}
\centering
\includegraphics[width=1\textwidth]{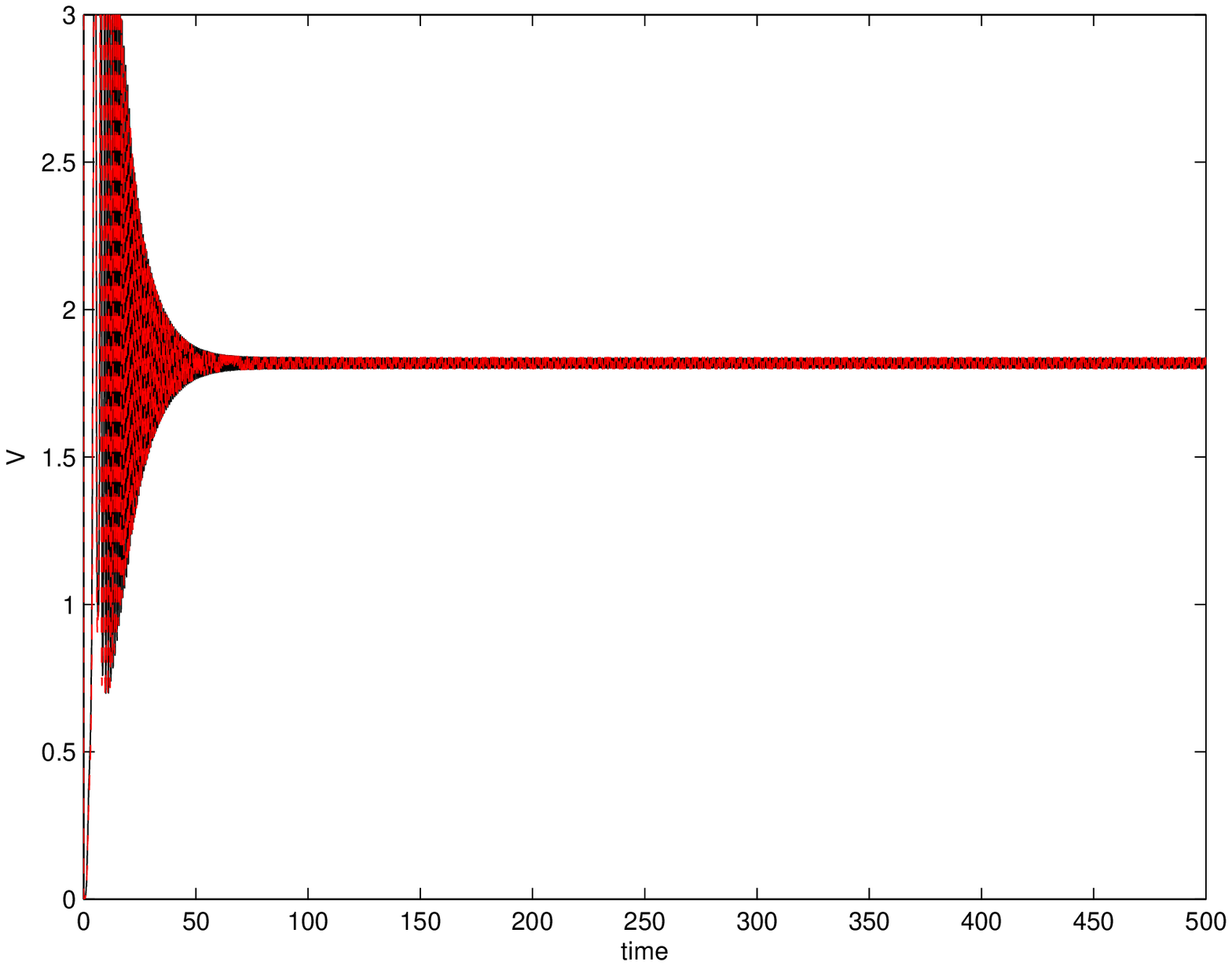}
\end{minipage}
\begin{minipage}[c]{.40\textwidth}
\centering
\includegraphics[width=1\textwidth]{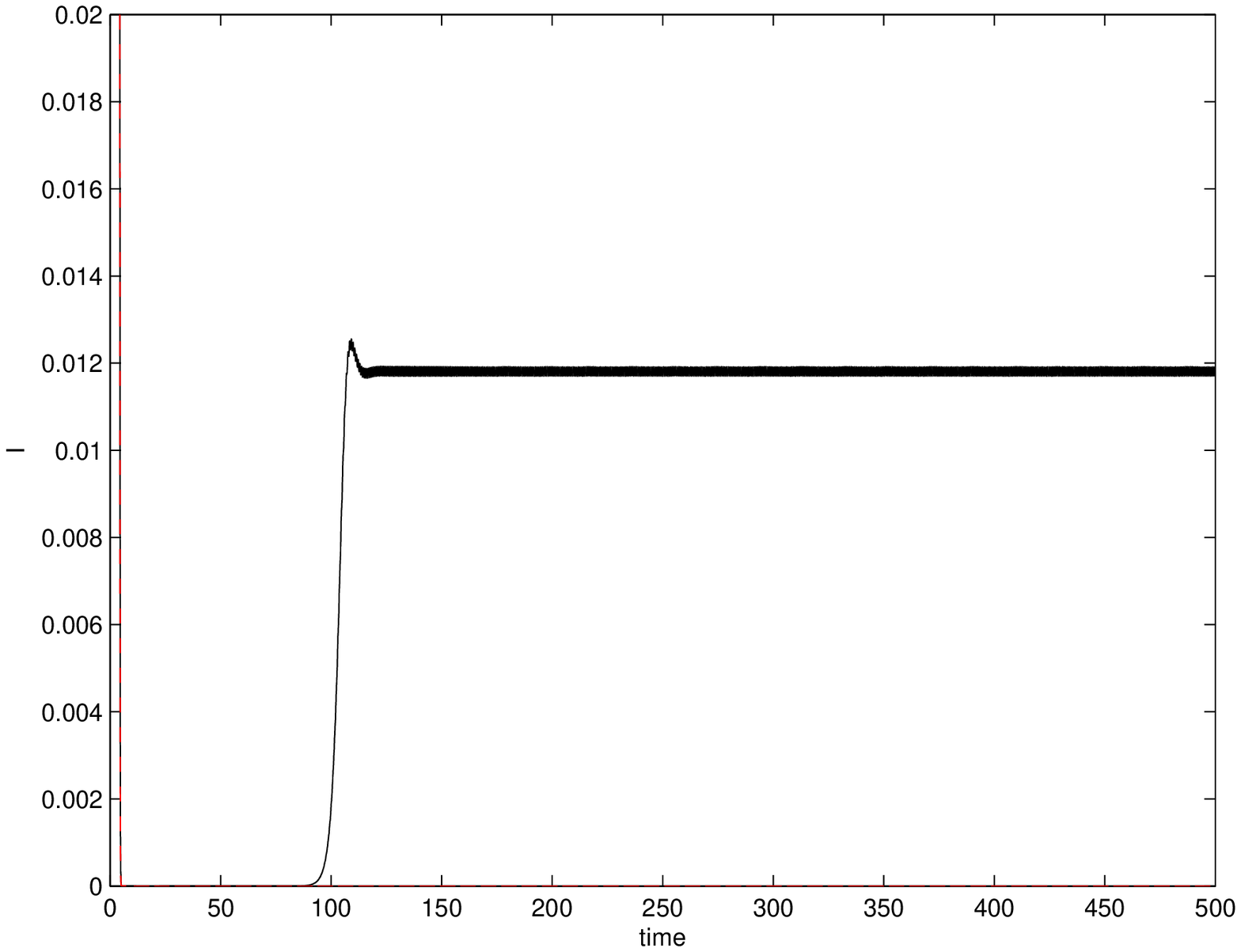}
\end{minipage}
\hspace{10mm}
\begin{minipage}[c]{.40\textwidth}
\centering
\includegraphics[width=1\textwidth]{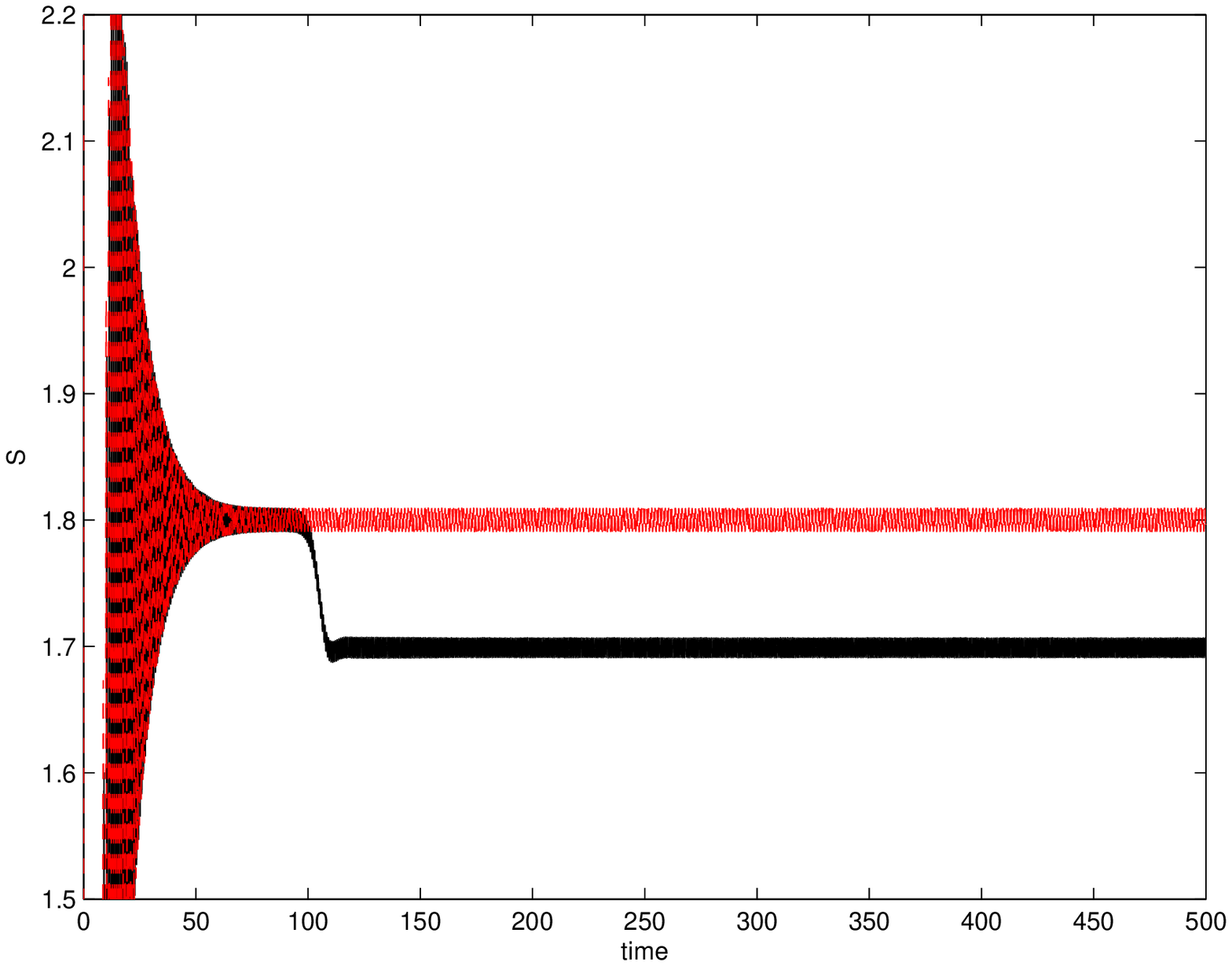}
\end{minipage}
\caption{The parameters are
$m = 20$, $p = 11$, $l = .4$, $ e = 1.8$, $ h = 11.5$, $ q = .5$, $ \beta = 4.5$,
$\gamma = .5$, $ \nu = .6$, $ a = 5$, $ K = 9$, $c = 2.2$.
The choice $n = 4$ leads to the equilibrium $\mathbf{E}_3$ (red), for  $n=3.6$ we find instead $\mathbf{E}^*$ (black).}
\label{fig_k}
\end{figure}

A further transcritical bifurcation is shown numerically in Figure \ref{fig_k}
for the parameter values
$m = 20$, $p = 11$, $l = .4$, $ e = 1.8$, $ h = 11.5$, $ q = .5$, $ \beta = 4.5$,
$\gamma = .5$, $ \nu = .6$, $ a = 5$, $ K = 9$, $c = 2.2$.
The choice $n = 4$ leads to the equilibrium $\mathbf{E}_3$, for  $n = 3.6$ we find instead $\mathbf{E}^*$.
The transcritical bifurcation occurs for $n^*=3.85$.

\section{Discussion}

Food chain models are now classical in the literature. Here, however, we have made a step further in that we allow epidemics to
affect one population in the chain.

The proposed ecoepidemic food chain presents some novel features that distinguish it from its disease-free counterpart.
The purely demographic model indeed exhibits a series of transitions for which the intermediate population emanates
from the situation in which only the lowest trophic level thrives
when the threshold condition (\ref{Qtilde_feas}) holds.
The top predator can invade this two-population situation again when
a second threshold is crossed, (\ref{Q*_feas}).
In all these cases there is only one stable equilibrium, which is globally asymptotically stable, and no Hopf bifurcations can arise.

Instead, the ecoepidemic food chain shows a much richer behavior in several ways.

At first,
there are more transcritical bifurcations: all the ones that appear already in the classical case show up here as well,
but furthermore there are new ones. In fact, for instance,
the healthy prey-only equilibrium can give rise to the endemic disease prey-only equilibrium,
if the disease contact rate exceeds a certain value. This can be recast as saying that the prey carrying capacity must be larger than
the ratio of the rates at which individuals leave and enter the infected class, i.e. the ratio of the sum of the recovery and mortality
rates over the disease contact rate.

Secondly,
in addition, it contains persistent limit cycles for some or all the populations thriving in it.
This occurs in spite of the very simple formulation of the equations. In fact, we do not assume anything else apart from
logistic growth for the populations, when applicable, i.e. at the lowest trophic level, and quadratic, or bilinear, interactions to
describe the interaction terms in the predation as well as in the disease transmission. No more sophisticated mechanisms such as
Holling type II terms or more complicated nonlinearities are present in the model.

Finally, bistability is discovered
among some equilibria, leading to situations in which the computation of their basins of attraction is relevant for
the system outcome in terms of its biological implications.

In one case there could be both the equilibria with only the bottom prey with an endemic disease, or the last two trophic levels can coexist
in a disease-free environment, compare respectively
the pair of equilibria $\mathbf{E}_5$ and $\mathbf{E}_4$. Evidently, if the epidemiologists want to fight
the disease, the latter is the goal to achieve. It represents also a good result from the biodiversity point of view, since in it two populations
of the chain survive instead of only one. Evidently, then, it is important to compare the basins of attraction of the two points and understand
how the system parameters do influence them. The goal would then be to act on these parameters in order to reduce to the minimum possible the
basin of attraction of the unwanted equilibrium point, in this case the one with the endemic disease.

A similar situation occurs between the point with only the bottom prey with endemic disease
$\mathbf{E}_5$ and the one containing all the chain's populations,
but disease-free, i.e. $\mathbf{E}_3$. Evidently, the latter represents even a better situation from the conservation and biodiversity point of
view. Under this perspective in the applied ecologist frame of mind,
the coexistence of all populations including the diseased individuals, and the top predators-free environment with
endemic disease represents a secondary choice with respect to the former bistability situation,
because in the former there is the disease-free equilibrium with all trophic levels thriving.
The influence of the bottom prey carrying capacity in the shaping of the bistable equilibrium coexisting with the top predator and disease-free
equilibrium, $\mathbf{E}_4$ has been investigated numerically. For a low value of $K$, the former equilibrium coexists stably with the one
in which only the bottom trophic level is present, with endemic disease. In this alternative the latter equilibrium represents a worse situation.
For larger values of the carrying capacity $\mathbf{E}_4$ coexists instead with the top predators-free environment. Again the latter introduces
the disease and therefore it should be regarded as a bad situation, but this kind of coexistence is preferable to the one we get for
lower $K$, since two trophic levels are in any case preserved, whether with or without the disease.

%\newpage

Comparing the food chain model to the pure SIS epidemic model,
we observe a much richer behavior in the former, since the latter exhibits only one equilibrium
at the time, in view of the existence of the transcritical bifurcations discussed in Section \ref{SIS}.
In addition this equilibrium, whether disease-free or endemic, is always stable,
as stated in Proposition 4. Therefore, the more
complex structure of the food chain entails the presence of persistent oscillations.

\begin{figure}[htbp]
\centering
\includegraphics[width=.5\textwidth]{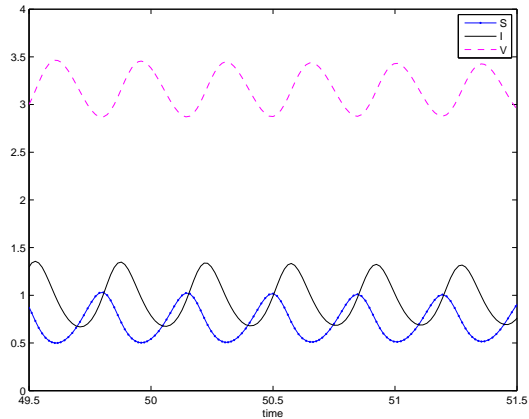}
\caption{$m = 0;p = 0;l = 5; e = .2; h = 0; q = 5; \beta = 19.5; n = 4; \gamma = .2; \nu = 1.5; a = 20; K = 350;c = .4;$}
\label{fig_relaz_c}
\end{figure}

To better investigate the subsystem with only the two lowest trophic levels behavior, we consider the coexistence
situation in the former, when all the populations thrive via sustained oscillations, as shown in
Figure \ref{fig_d_new}.
If we keep the same parameter values, but disregard the top predator $W$, setting also $m=p=h=0$,
the ecoepidemic subsystem settles to a {\textbf{stable}} equilibrium. On the other hand, we can start from the
unstable situation in the ecoepidemic subsystem, which can be obtained if
any condition in (\ref{ecoepid_st}) does not hold.
For the parameters $l = 5$, $e = 0.2$, $q = 5$, $\beta = 19.5$,
$n = 4$, $\gamma = 0.2$, $\nu = 1.5$, $a = 20$, $K = 350$, $c = 0.4$, the unstable behavior is
shown in Figure \ref{fig_relaz_c}.
In this case, introducing now the new population $W$, we find that the system settles to an
endemic equilibrium
in which only the infected lower trophic level population survives, equilibrium $\mathbf{E}_5$.
This occurs for the parameter
values $m = 0.2$, $p = 0.2$, $h = 0.8$.
A second example leading from persistent oscillations in the ecoepidemic subsystem to stable coexistence
in the full model is obtained
instead for the parameter choice 
$m = 0.3$, $p = 0.2$, $l = 5$, $e = 0.1$, $h = 0.02$, $q = 5$, $\beta = 19.5$, $n = 4$, $\gamma = 0.2$,
$\nu = 1.5$, $a = 20$, $K = 350$, $c = 0.4$. We obtain in this case the stable coexistence equilibrium with
the following population
values, $\mathbf{E}^* = (6.267, 1.504, 1.027, 0.390)$.
These results show that in the ecoepidemic food chain model and in the ecoepidemic subsystem
their two respective coexistence equilibria are independent of each other.

%\newpage

\bibliography{references}
\bibliographystyle{chicago}

\end{document}